\documentclass[multphys,vecphys]{svmult}

\usepackage{makeidx}     
\usepackage{graphicx}    
\usepackage{multicol}    
\makeindex             

\newcommand{\tauc}{\tau_{\scriptscriptstyle T}}
\newcommand{\tauh}{\tau_{\scriptscriptstyle H}}
\newcommand{\taumin}{\tau_{\rm min}}
\newcommand{\rhob}{\overline{\rho}}
\newcommand{\kat}{\kappa_{\scriptscriptstyle T}}
\newcommand{\leqa}{\stackrel{<}{\scriptstyle \sim}}
\newcommand{\geqa}{\stackrel{>}{\scriptstyle \sim}}

\newcommand{\kfa}{\bar{k}_{\scriptscriptstyle F}}
\newcommand{\kfr}{\bar{k}_{\scriptscriptstyle F} R}
\newcommand{\kb}{k_{\scriptscriptstyle B}}
\newcommand{\ef}{E_{\scriptscriptstyle F}}
\newcommand{\efa}{\overline{E}_{\scriptscriptstyle F}}
\newcommand{\efo}{\widetilde{E}_{\scriptscriptstyle F}}
\newcommand{\epsf}{\epsilon_{\scriptscriptstyle F}}

\newcommand{\ba}{{\overline {\cal B}}}
\newcommand{\bt}{{\widetilde {\cal B}}}

\begin{document}

\title*{Regularity and chaos in the nuclear masses} 
\author{Patricio Leboeuf}
\institute{Laboratoire de Physique Th\'eorique et Mod\`eles
Statistiques,\footnote[1]{Unit\'e de Recherche de l'Universit\'e Paris XI
associ\'ee au CNRS} \\ B\^at. 100, Universit\'e de Paris-Sud, 91405 Orsay
Cedex, France
\abstract{Shell effects in atomic nuclei are a quantum mechanical
manifestation of the single--particle motion of the nucleons. They are
directly related to the structure and fluctuations of the single--particle
spectrum. Our understanding of these fluctuations and of their connections
with the regular or chaotic nature of the nucleonic motion has greatly
increased in the last decades. In the first part of these lectures these
advances, based on random matrix theories and semiclassical methods, are
briefly reviewed. Their consequences on the thermodynamic properties of Fermi
gases and, in particular, on the masses of atomic nuclei are then presented.
The structure and importance of shell effects in the nuclear masses with
regular and chaotic nucleonic motion are analyzed theoretically, and the
results are compared to experimental data. We clearly display experimental
evidence of both types of motion.}
\texttt{leboeuf@lptms.u-psud.fr} \\
\vspace{0.5cm}
Lectures delivered at the VIII Hispalensis International Summer School,
Sevilla, Spain, June 2003 (to appear in {\sl Lecture Notes in Physics}, 
Springer--Verlag, Eds. J. M. Arias and M. Lozano).}
\maketitle

\section{Introduction}
\label{sec:1}
\index{Introduction}

The mass is one of the most basic properties of an atomic nucleus. In spite of
its fundamental character, it is a quite complicated and highly non--trivial
quantity. According to Einstein's celebrated law, $m = E/c^2$, all types of
energy stored inside a nucleus contribute to its mass. If we imagine that the
nucleons are initially separated, and then progressively move towards each
other to finally form a nucleus in its ground state, a certain amount of
energy ${\cal B}$ will be released in the process. Mathematically, this
corresponds to the equation,
\begin{equation} \label{defb}
M c^2 = \sum_j m_j \ c^2 - {\cal B} \ ,
\end{equation}
where $M$ is the mass of the nucleus, $c$ the speed of light, and $m_j$ the
mass of the $j$th nucleon. The binding energy ${\cal B}$ is responsible for
the cohesion of the system, the larger ${\cal B}$ is, the more stable is the
nucleus. There are different sources to this energy. The most important one
comes from the strong attractive interaction between nucleons. Other
interactions and effects that contribute to ${\cal B}$ are the Coulomb
repulsion between protons, the surface effects, etc. In phenomenological
descriptions of the mass, these contributions are usually taken into account
by, for example, liquid drop expressions \`a la von Weizs\"acker \cite{wei}.
There is another important contribution. In a semiclassical picture where each
nucleon is thought of as an individual particle moving in a mean--field
potential, this contribution is related to the motion of the nucleons inside
the nucleus. Because of Pauli's exclusion principle, when the nucleons are put
together to form a bound system they are not at rest, but instead move with a
speed which is of the order of $0.3 c$. Via Einstein's relation, this kinetic
energy also contributes to ${\cal B}$, and therefore to the mass. Part of this
energy, namely the one that varies smoothly with the number of nucleons, is
already taken into account in the phenomenological liquid drop terms mentioned
above. In contrast, the remaining part of the kinematic energy fluctuates with
the number of nucleons. It moreover depends on the nature of the motion of the
nucleons. For this reason, we refer to this contribution as the ``dynamical''
component of the mass.

From the theory of dynamical systems, there are two important, extreme, and
distinct possibilities for the motion of the nucleons inside the nucleus,
namely it can be either regular or chaotic. A natural question to ask concerns
the relation between these two types of motion and the properties of the
associated dynamical mass. Is it possible to distinguish between a ``regular''
and a ``chaotic'' mass? How can the difference be detected experimentally, if
any? The purpose of these lectures is to bring some answers in this direction.
We will show that it is indeed possible to make the difference between both
types of dynamical masses via their fluctuation properties, i.e., how the mass
varies as a parameter, typically the number of nucleons, is changed. When the
results are applied to the available experimental data, the analysis of both
types of dynamical masses indicates that, apart from a dominant regular
component that will be described in some detail via a spheroidal potential,
there might be in addition some chaos in the ground state of the nucleus. The
corresponding contribution to the mass and its dependence with the number of
nucleons will be explicitly computed. It will be shown that the hypothesis of
chaotic layers explains the observed differences between the experimental data
and previous theoretical calculations.

Although the setting we are using is based on a single--particle picture of
the nucleus, several arguments indicate that the results obtained are of more
general validity. They may in fact be related to correlations acting between
nucleons that are beyond a mean--field picture.

Traditionally, dynamical effects in the structure of nuclei are referred to as
shell effects. This designation originated in atomic physics, where the
symmetries of the Hamiltonian produce strong degeneracies of the electronic
levels (the shells). These degeneracies induce, in turn, oscillations in the
electronic binding energy. Shell effects are therefore due to deviations (or
bunching) of the single--particle levels with respect to their average
properties. The degeneracies of the electronic levels of an atom produced by
the rotational symmetry are an extreme manifestation of level bunching. In
general, in systems that have other symmetries, or have no symmetries at all,
there will still be level bunching, but its importance will typically be
minor. Therefore, depending on the presence or absence of symmetries, the
shell effects may be more or less important. The level bunching, and more
generally the fluctuations of the single--particle energy levels, are thus a
very general phenomenon. The theories that describe those fluctuations make a
neat distinction between systems with different underlying classical dynamics
(e.g., regular or chaotic). It is therefore our purpose, before entering in
the discussion of the nuclear masses in \S \ref{sec:5}, to give a general
(though elementary) presentation of the fluctuation properties of the level
density (\S \ref{sec:2} and \ref{sec:3}), and of the way these fluctuations
manifest in the different thermodynamic properties of a Fermi gas (\S
\ref{sec:4}). It will be shown, in particular, that generically the size of
the fluctuations is more important when the classical underlying motion is
regular, as compared to the chaotic case. This fact may seem paradoxical,
since we are usually willing to associate chaos with noise and fluctuations.
However, concerning the shell effects in quantum mechanics, the situation is
exactly the other way around: because of the instability of the classical
orbits, level clustering is less important in chaotic systems. We will
precisely quantify the difference for several thermodynamic quantities. We
will moreover discuss universality, e.g. the validity in large classes of
systems of common features of the fluctuations.

The ``dynamical'' or shell fluctuations of the mass \cite{bm,sm} considered
here are a particular example of a general phenomenon. Similar fluctuations
are expected to occur, with different degrees of importance, in all
thermodynamic quantities of a fermionic gas. This point is discussed in some
detail in \S \ref{sec:4}. Many illustrative examples may be mentioned, like
the fluctuations of the persistent currents in mesoscopic rings, the force
fluctuations observed when pulling a metallic nanowire, or the supershell
structure in metallic particles. They all have the same physical origin, which
in the context of semiclassical theories is associated to classical periodic
orbits. See for example Ref.~\cite{imry} for a discussion in condensed matter
mesoscopic physics, and Ref.~\cite{brack} for cluster physics. For a general
statistical theory of the fluctuations, and further related references, see
Ref.~\cite{lm3}.

These lectures are based on research carried out over the past few years,
during which I have enjoyed and benefitted from collaborations with A.
Monastra and O. Bohigas. I am deeply indebted to them. \S \ref{sec:4} is based
on Leboeuf and Monastra, Ref.~\cite{lm3}, and \S \ref{sec:5} on Bohigas and
Leboeuf, Ref.~\cite{bl}. The analysis of the regular component of the mass and
of supershell structures in nuclei presented in \S \ref{sec:5}, based on a
spheroidal cavity with a finite lifetime for quasiparticles, is original and
was not published elsewhere. References \cite{lmb} and \cite{lm4} report some
closely related results that are not discussed here; they treat thermodynamic
aspects of a fictitious element, ``The Riemannium'', a schematic many body
system inspired from number theory.

\section{Local Fluctuations: Random Sequences}
\label{sec:2}

Consider a bound single--particle Hamiltonian $H$, whose quantum mechanical
spectrum is given by a discrete sequence of energy levels $\{ E_1, E_2,
\ldots, E_i, \ldots \}$. $H$ may either represent a self--consistent mean
field approximation of an interacting system, or simply the Hamiltonian of a
given one--body problem. At a given energy $E$, we denote the typical distance
between neighboring eigenvalues, the mean level spacing, by $\delta$. The aim
in this section is to present a short overview of some of the results that
have been obtained concerning the fluctuations of the single--particle energy
levels on scales of order $\delta$, and to establish connections with the
underlying classical dynamics. Fluctuations on a scale $\delta$ are termed
``local'', compared to fluctuations on larger scales to be discussed in \S
\ref{sec:3}.

One of the most pervasive theories in the description of the statistical
properties of sequences of numbers is the random matrix theory (RMT). Its
range of applicability largely exceeds the spectra of single--particle
systems, and even the frontiers of physics. For example, one can find it in
the description of nuclear, atomic and molecular systems, the motion of
electrons in a disordered potential, the behavior of classically chaotic
systems, the study of integrable models, the description of the statistical
properties of the critical zeros of the Riemann zeta function, and in problems
of combinatorics. Some interesting articles and lectures covering these topics
can be found in Refs.~\cite{porter}--\cite{jpa}.

Motivated by its mathematical simplicity, one of the original ensembles
introduced is the so-called Gaussian ensemble of random matrices
\cite{wigner}, defined as the set of hermitian matrices $H$ whose elements are
Gaussian independent variables. The probability density of a given $N \times
N$ hermitian matrix is defined as
\begin{equation} \label{1}
P(H)= c_H \exp\left(-\frac{\beta}{2} {\rm tr} H^2 \right),
\end{equation} 
where $c_H$ is a normalization constant. A very specific aspect of this
ensemble is its invariance under rotations in Hilbert space. The form of $H$
depends on symmetry considerations: its matrix elements are real for even spin
systems with time reversal symmetry, complex in the absence of time reversal
symmetry, and quaternion for odd spins with time reversal symmetry. For these
three cases, the parameter $\beta$ in Eq.~(\ref{1}) takes the value $1$, $2$
and $4$, respectively.

Given the probability density of matrix elements (\ref{1}), the problem is to
find the probability distribution of the associated eigenvalues $E_i$,
$i=1,\ldots,N$ and eigenvectors $\{p_k\}$ of $H$ (the number of parameters in
the parametrization $p_k$ of the eigenvectors depends on the symmetry of $H$).
In the basis where $H$ is diagonal, ${\rm tr} H^2 = \sum\limits_{i} E_i^2$.
Moreover, the Jacobian of the transformation from matrix elements to
eigenvalues and eigenvectors is given by \cite{mehta}
\begin{equation} \label{2}
{\cal J}=\frac{\partial (H_{ij})}{\partial (E_\alpha,p_k)}=\prod_{i<j=1}^N 
|E_i - E_j|^\beta \ .
\end{equation}
The eigenvector's components are absent in this expression because of the
rotational invariance of the ensembles. From these results, it follows that
the joint probability density of the eigenvalues is (in an abuse of notation,
we use the same symbol $P$ as for the matrix probability density),
\begin{equation} \label{3}
P(E_1,\ldots,E_N) = c_E \prod_{i<j=1}^N |E_i - E_j|^\beta \exp\left(-
\frac{\beta}{2} \sum_{i=1}^N E_i^2 \right) \ ,
\end{equation}
where $c_E$ is a normalization constant.

The most characteristic feature of this probability density is the existence
of a strong repulsion between eigenvalues. The repulsion may be seen, in
Eq.~(\ref{3}), from the fact that $P \rightarrow 0$ when $E_j \rightarrow
E_i$. Being related to the Jacobian, this repulsion is not a particular
feature of the ensemble (\ref{1}), but rather constitutes a very basic and
general fact of most quantum mechanical systems.

From Eq.~(\ref{3}) the $k$--point correlation function, 
$$
R_k (E_1,\ldots,E_k)=N!/(N-k)! \int\ldots\int dE_{k+1} \ldots dE_N
P(E_1,\ldots,E_N) \ ,
$$
may be explicitly computed. We are particularly interested in the case $k=2$,
that defines the density of pairs of eigenvalues separated by a distance $y =
E_2 - E_1$ (when the distance $y$ is local (i.e., on a scale $\delta$), the
two--point function is stationary and depends only on the difference;
moreover, we are here interested in the ``bulk'' results, i.e. statistics of
eigenvalues located near the center of the spectrum, far from the edges). A
convenient way to characterize the function $R_2 (y)$ is through its Fourier
transform,
\begin{equation} \label{5}
K (\tau) = 2 h \int_{0}^{\infty} \cos (y \tau /\hbar ) \left[ R_2
(y) - \rhob^2 + \rhob \ \delta (y) \right] \ d y \ ,
\end{equation}
usually called the form factor. $K (\tau)$ has units of time; $\rhob$ is the
average density of states,
\begin{equation} \label{rhob0}
\rhob = \delta^{-1} = \tauh / h \ ,
\end{equation}
where we have introduced the Heisenberg time, the conjugate time to the mean
level spacing $\delta$. From Eq.~(\ref{3}), the random matrix form factor is
found to be \cite{mehta},
\begin{equation} \label{krmt}
K_{rmt} (\tau) = \left\{ \begin{array}{ll} 
\left[ 2 \tau - \tau \log \left( 1 + \frac{2 \tau}{\tauh} \right) \right] 
\Theta (\tauh - \tau) \ + \\
\ \left[ 2 \tauh - \tau \log \left( \frac{2 \tau + \tauh}{2 \tau - \tauh} \right)\right] 
\Theta (\tau - \tauh)  \;\;\;\;\;\;\;\;\;\;\;\;\;\;\; & \beta = 1 \\ \\
\tau \ \Theta (\tauh - \tau) + \tauh \ \Theta (\tau - \tauh)  & \beta = 2 
\ . \\
\end{array}
\right.
\end{equation}
For simplicity, we have restricted to $\beta=1$ and $2$ (since the spin
dynamics will not be included in the following, we do not consider the
symplectic symmetry). In (\ref{krmt}) the function $\Theta$ is Heaviside's
step function. $\tauh$ is the only characteristic time scale in
Eq.~(\ref{krmt}). For short times $\tau \ll \tauh$, the form factor behaves
as,
\begin{equation} \label{krmtd}
K_{rmt} (\tau) = \frac{2}{\beta} \ \tau \ , \;\;\;\;\;\;\;\;\; \tau \ll \tauh
\ ,
\end{equation}
whereas for $\tau \gg \tauh$ it tends to $\tauh$ (this is a general property
of any discrete sequence of levels).

Another useful quantity is the nearest-neighbor spacing distribution $p(s)$,
defined as the probability to find two neighboring eigenvalues separated by a
distance $s$, where $s$ is here measured in units of the mean level spacing
$\delta$. Though the limiting behavior $N \rightarrow \infty$ of $p(s)$ may be
computed analytically, the result is not explicit and it is customary to use
the expression obtained from $2 \times 2$ matrices (which turns out to differ,
by accident, by only a few percent from the exact expression),
\begin{equation} \label{6}
p(s) = \left\{ \begin{array}{ll} \frac{\pi}{2} s \exp\left( -\frac{\pi}{4} s^2
\right) \;\;\;\;\;\;\;\;\;\;\;\;\;\;\; & \beta = 1 \\ \\ \frac{32}{\pi^2} s^2
\exp\left( -\frac{4}{\pi} s^2 \right) & \beta = 2 \ . \\
\end{array}
\right.
\end{equation}

The results obtained from RMT are usually compared to an individual system,
whose spectrum has been computed numerically or measured experimentally. The
statistical properties of the system are calculated in some window of size
$\Delta E$ located around an energy $E$. To obtain statistically significant
results, an average is done by changing the position of the window. The window
size $\Delta E$ should be sufficiently large in order to include many
single--particle levels and make a statistical analysis meaningful, but also
it should not be too large, to guarantee that the gross features of the
spectrum (e.g., the mean level spacing) are constant.

\begin{figure}
\centering
\includegraphics[height=5cm]{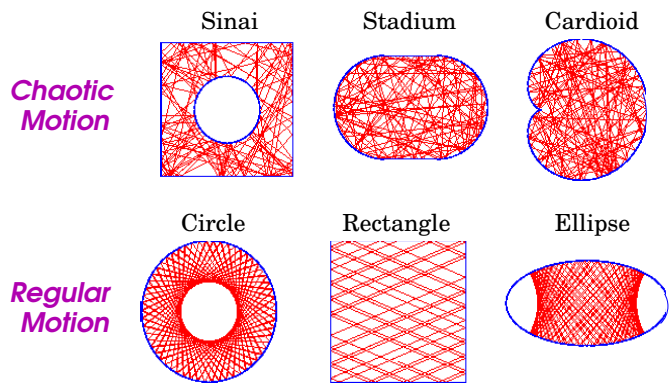}
\caption{Examples of classical motion inside 2D cavities. Upper part: chaotic,
lower part: integrable. In each cavity, a typical trajectory is shown. Taken
from \cite{ab}.}
\label{fig:1}       
\end{figure}

\begin{figure}
\centering
\includegraphics[height=6cm]{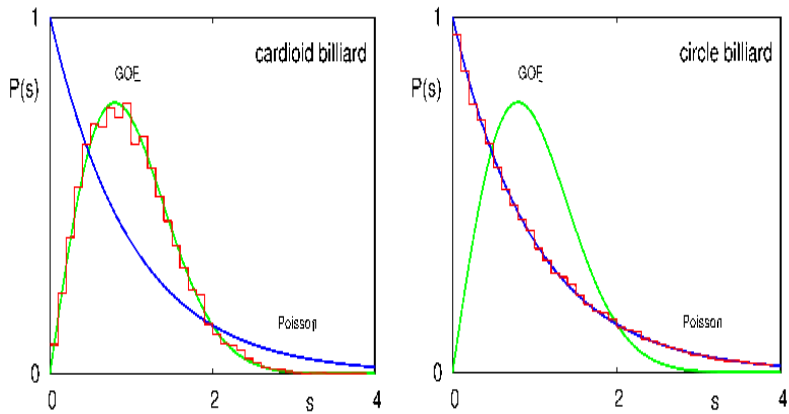}
\caption{Nearest neighbor spacing distribution for the cardioid and circle
billiards, compared to RMT ($\beta=1$) and Poisson distributions. 
Taken from \cite{ab}.}
\label{fig:2}       
\end{figure}

The RMT, as formulated above, is a one parameter theory. The results (see
Eq.~(\ref{krmt})) depend only on the mean level spacing $\delta$ (or
Heisenberg time $\tauh$). In order to compare different spectra coming from
different physical systems, the mean spacing should be normalized. Doing this,
good agreement with RMT of the local fluctuations has been found in many
different situations. This is the case for example of the neutron resonances
in heavy nuclei, which was the original experimental motivation to develop the
RMT. Much simpler though non--trivial problems were subsequently analyzed. At
the beginning of the eighties it was shown that, in the metallic regime, the
statistical properties of the eigenvalues of the famous Anderson problem
(e.g., an electron moving in a metal with impurities, modeled by a random 3D
potential) coincide with RMT \cite{efetov}. Up to now, this is the only case
were the agreement with RMT was proved rigorously, though the model is not
fully deterministic (randomness is incorporated by hand in the Hamiltonian).
In fully deterministic systems, RM statistics were conjectured to hold
generically for high--lying eigenvalues of systems with a classical fully
chaotic motion. This is the Bohigas--Giannoni--Schmit (BGS) conjecture in
quantum chaos \cite{bgs,bohigas}, supported by many numerical results and by
some analytic arguments (see next section). One of the simplest Hamiltonians
allowing to test the conjecture, and that will be here employed as a model
system, is to consider the motion of a free particle inside a cavity with
perfect elastic reflections on the boundary. The upper part of
Fig.~\ref{fig:1} shows several fully chaotic two--dimensional (2D) examples:
the Sinai billiard, the stadium, and the cardioid. The quantization of the
motion is given, for a billiard system, by the Schr\"odinger equation of a
free particle with Dirichlet boundary conditions imposed (i.e., the wave
function should vanish on the border). In the left part of Fig.~\ref{fig:2} is
shown the nearest neighbor spacing distribution of high--lying eigenvalues of
the cardioid billiard obtained numerically. The cardioid billiard is
time--reversal symmetric, and is therefore compared with the ensemble $\beta =
1$ in Eq.~(\ref{6}) \cite{foot1}.

A relevant question is the universality of the ensemble of matrices
considered. There is no special reason, aside technical ones, to consider the
particular Gaussian distribution of matrix elements described above. Other
statistical weights could have been chosen as well. It is therefore important
to understand to which extent the results are independent of the distribution.
Some form of universality of the local statistics has been shown. As the size
of the matrix increases, the bulk eigenvalue distribution converges towards a
limit that, to a large extent, is independent of the probability measure used
\cite{bz,sosh}.

Through the BGS conjecture, RMT is associated to the concept of complexity,
here manifested as chaotic classical motion. In contrast, there are large
classes of physical systems were the conjecture does not apply. An extreme and
well known class is that of the regular, or integrable, systems. These are
systems with as many independent integrals of motion as degrees of freedom.
The lower part of Fig.~\ref{fig:1} shows several two--dimensional (2D)
examples. In the circular billiard, for instance, the two independent
integrals of motion are the energy and the angular momentum. The existence of
these constants of the motion forces the orbits to form regular patterns, and
no exponential sensitivity to perturbations, like in chaotic systems, exists.
Regular and fully chaotic motion are two extreme situations. A billiard with
an arbitrary shape will generically contain both types of dynamics, regular
and chaotic (mixed system). Some aspects of mixed dynamics will be discussed
in \S \ref{5}.

When a regular system is quantized, the statistical analysis of the
fluctuations give results that are quite different with respect to RMT. What
is found is that the local statistics of the corresponding quantum energy
levels are described by a Poisson process. (e.g., the eigenvalues behave,
statistically, as a set of uncorrelated levels) \cite{foot2}. Many numerical
simulations support this conjecture, though there is no general proof, aside
some particular systems \cite{marklof}. This property is illustrated on
the right hand side of Fig.~\ref{fig:2}, where the nearest neighbor spacing
distribution of a circular 2D billiard is compared to the Poisson result,
$p(s) = {\rm e}^{-s}$ ($\delta = 1$). The form factor of an uncorrelated
sequence of levels is independent of $\tau$,
\begin{equation}\label{kint}
K_{unc}(\tau) = \tauh \ .
\end{equation} 

In summary, and loosely speaking because there are exceptions that we don't
discuss here, for the two types of dynamics considered the local correlations
of energy levels are universal and either described by random matrix theory in
chaotic or ``complex'' systems, or by an uncorrelated Poisson sequence for
integrable or ``regular'' systems.

\section{Long Range Fluctuations: Semiclassics}
\label{sec:3}

In fact, and contrary to what Fig.~\ref{fig:2} suggests, RMT or a Poisson
process do not describe faithfully the fluctuation properties of real physical
systems. The reason is that these simple ensembles fail to describe some
important physical features related to the short time dynamics. There is only
one time scale that can be associated, through quantum mechanics, to a
stationary sequence of random numbers. This is the Heisenberg time, that
corresponds to the conjugate time to the mean level spacing. But physical
system possess other time scales, that are typically smaller than $\tauh$. In
ballistic systems (e.g., no disorder present), like a particle moving freely
inside a 2D or 3D cavity, another important time scale is the time of flight
across the system, denoted $\taumin$. We will give a precise definition of
$\taumin$ below using the periodic orbits. This is typically the smallest
relevant time scale of a dynamical system. It corresponds to a large energy
scale, given by,
\begin{equation}\label{ec}
E_c = h / \taumin \ .
\end{equation} 
We will now see that the typical single--particle energy levels of a ballistic
dynamical system have, superimposed to the universal local fluctuations on a
scale $\delta$ discussed in the previous section, system--specific long--range
modulations on a scale $E_c$. These modulations of the density with respect to
the average properties of the spectrum are not described by random sequences
of numbers (correlated or not).

It turns out that a very convenient tool to analyze in a systematic way the
deviations of a quantum spectrum with respect to its average properties is the
semiclassical analysis. This method is convenient from either a technical as
well as a conceptual point of view, and is justified if a classical limit of
the quantum problem exists. This limit can be obtained by different
procedures. For example, in the context of nuclear physics it is known that a
classical limit of some simple shell models of the nucleus can be obtained by
letting the number of particles tend to infinity \cite{koonin,ls}. More
generally, the connexion with a classical dynamics can be done through the
mean field, which reduces the fully interacting many--body problem to a
single--particle motion in a self--consistent potential. The interest of
semiclassical theories is that they offer a unified scheme of analysis, and
provide a conceptual framework that allows in many cases to understand a
physical property using simple ideas.

The basic object we are interested in is the single--particle density of
states
\begin{equation} \label{rho}
\rho (E,x) = g_s \sum_j \delta \left[ E - E_j(x) \right] \ .
\end{equation}
The $E_j(x)$ are the discrete eigenvalues of the single--particle Hamiltonian
$H$. The eigenvalues depend on a set of parameters $x$ that fix, for example,
the shape of the potential, or may represent any other external parameter. The
prefactor $g_s$ accounts for spin degeneracy.

The classical motion is described by a function $H (\vec{p},\vec{q})$ of the
phase--space variables $(\vec{p},\vec{q})$. In the semiclassical limit $\hbar
\rightarrow 0$ the quantum density of states can be approximated by a sum of
smooth plus oscillatory terms,
\begin{equation} \label{rhos}
{\rho} (E,x) = g_s \left[ \overline{{\rho}} (E,x) + \widetilde{{\rho}} (E,x) 
\right] \ .
\end{equation}
The first term is a smooth function of $E$ that describes the average
properties of the spectrum. It is usually known as the Thomas--Fermi
contribution (or Weyl expansion for the motion of a particle inside a
hard--wall cavity). The leading term of $\overline{{\rho}}$ is given by the
well-known semiclassical rule that associates, in a $D$--dimensional system, a
phase--space volume $h^D$ to each quantum state,
\begin{equation} \label{rhob}
\overline{{\rho}} = \frac{1}{(2 \pi \hbar)^D} \int d^D p \ d^D q \
\delta (E - H (\vec{p},\vec{q})) \ . 
\end{equation}
Corrections to Eq.~(\ref{rhob}) depend on derivatives of $H$ \cite{brackbook}.
In the case of the motion inside a $3D$ cavity, the expansion takes a
particularly simple form \cite{bh},
\begin{equation}\label{rhobc}
{\bar \rho} = \frac{V}{4 \pi^2} \left( \frac{2 m}{\hbar^2}
\right)^{3/2} E^{1/2} - \frac{S}{16 \pi} \frac{2 m}{\hbar^2}
+ \frac{\cal L}{16 \pi} \left( \frac{2 m}{\hbar^2}
\right)^{1/2} E^{-1/2} + {\cal O} (E^{-3/2}) \ .
\end{equation}
Here $V$ and $S$ are the volume and the surface of the cavity, and ${\cal L}$
is a typical length that depends on its topology; $m$ is the mass of the
particle. The integral of the density with respect to the energy is also of
interest,
\begin{equation}\label{n}
{\cal N} (E,x) = g_s \sum_j \Theta \left[ E - E_j(x) \right] \ ;
\end{equation}
it gives the number of single--particle levels with energy $E_j \leq E$
(counting function). It
can be decomposed as in Eq.~(\ref{rhos}), 
\begin{equation}\label{nss}
{\cal N} (E,x) = g_s \left[{\overline {\cal N}} (E,x) + {\widetilde {\cal N}}
(E,x) \right] \ .
\end{equation}
The average part, obtained for a 3D cavity by integrating (\ref{rhobc}) with
respect to $E$, is
\begin{equation}\label{nb}
{\overline {\cal N}} =  \frac{V}{6 \pi^2} \left( \frac{2 m}{\hbar^2}
\right)^{3/2} E^{3/2} - \frac{S}{16 \pi} \frac{2 m}{\hbar^2} E
+ \frac{\cal L}{8 \pi} \left( \frac{2 m}{\hbar^2} \right)^{1/2} E^{1/2} 
+ {\cal O} (1) \ .
\end{equation}

Deviations with respect to the smooth behavior of the density are described by
the fluctuating part $\widetilde{{\rho}}$ in Eq.~(\ref{rhos}). They are given,
to leading order in an $\hbar$--expansion, by \cite{bb,gutz},
\begin{equation} \label{rhoo} 
\widetilde{{\rho}} (E,x)= 2 \sum_p \sum_{r=1}^{\infty} A_{p,r} (E,x) \cos
\left[ r S_p(E,x)/\hbar+\nu_{p,r} \right] \ . 
\end{equation}
The sum is over all the periodic orbits $p$ of the classical Hamiltonian $H
(\vec{p},\vec{q})$. The index $r$ takes into account multiple traversals
(repetitions) of a primitive orbit $p$. The orbits are characterized by their
action $S_p = \oint \vec{p} . d \vec{q}$, period $\tau_p=d S_p/dE$, stability
amplitude $A_{p,r}$, and Maslov index $\nu_{p,r}$. The functional form of
$A_{p,r}$ depends on the nature of the dynamics. In integrable system, the
periodic orbits form continuous families, with all members of a family having
the same properties. In contrast, in fully chaotic systems all orbits are
unstable and isolated. This important difference is at the origin of the fact,
to be discussed in the next sections, that shell corrections are more
important in integrable systems than in chaotic ones.

Note: {\scriptsize Why a sum over periodic orbits? This can be understood from
the following simplified argument. The density of states may be calculated
from the Green function, $\rho (E) = - \frac{1}{\pi} {\rm Im} \int d^D q \\
\lim_{\epsilon \rightarrow 0^+} G(\vec{q},\vec{q},E+ i \epsilon)$. In the
Feynman representation, $G(\vec{q},\vec{q},E)$ is the propagator for paths of
energy $E$ that start at $\vec{q}$ and come back to $\vec{q}$. In the limit
$\hbar \rightarrow 0$, the leading contribution to $G$ is a sum over all the
{\sl classical} trajectories that start and come back to $\vec{q}$. Finally, a
stationary phase approximation of the above integral over $\vec{q}$ selects,
among all the classical closed trajectories, those that start and come back to
a given point with the same momentum. Those are the periodic orbits}.

The fluctuating part $\widetilde{{\cal N}}$ of the counting function may be
obtained by integration of Eq.~(\ref{rhoo}) with respect to the energy. To
leading order in a semiclassical expansion, the integration with respect to
the rapidly varying factor $S_p(E,x)/\hbar$ dominates. This gives, to leading
order,
\begin{equation} \label{no}
\widetilde{{\cal N}} (E,x)= 2 \hbar \sum_p \sum_{r=1}^{\infty} 
\frac{A_{p,r}}{r \tau_p} \sin \left[ r S_p/\hbar+\nu_{p,r} \right] \ . 
\end{equation}

\begin{figure}
\centering
\includegraphics[height=5cm]{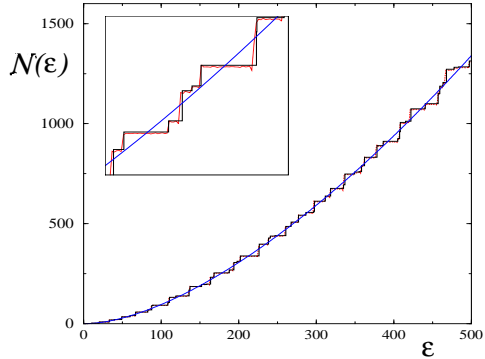}
\caption{Counting function of a spin--$1/2$ particle in a 3D spherical cavity.
Full staircase: quantum mechanical. Dotted staircase: semiclassical (almost
indistinguishable from the quantum mechanical one). Smooth interpolating
curve: Weyl's formula. Detail in the inset.}
\label{fig:3}       
\end{figure}

To illustrate the semiclassical approximation of the density of states and of
the counting function in a concrete example, we consider the free motion of a
spin--$1/2$ particle inside a three--dimensional (3D) cavity, elastically
reflected off the surface. The spherical cavity and the corresponding
semiclassical approximation have been extensively studied in the past, and I
refer the reader to some of those references for a more detailed presentation
\cite{bb,brackbook}. Quantum mechanically, the single--particle energy levels
$E_{n,l}$ of a free particle of mass $m$ moving inside a cavity of radius $R$
with Dirichlet boundary conditions are given by the quantization condition,
\begin{equation} \label{qs}
J_{l+1/2}\left(\sqrt{\epsilon_{n,l}} \right) = 0 \ , \;\;\;\;\; n=1,2,\ldots \
, \;\; l=0,1,2,\ldots \ ,
\end{equation}
where $E_{n,l} = \hbar^2 k_{n,l}^2 /2 m = \hbar^2 \epsilon_{n,l} /2 m R^2 =
E_0 \ \epsilon_{n,l}$, $J$ is the Bessel function, and $k$ the wavenumber. The
previous expressions define the quantities $\epsilon_{n,l}=(k_{n,l} R)^2$ and
$E_0=\hbar^2 /2 m R^2$. The index $l$ labels the angular momentum of the
particle, and $n$, that serves to identified the different zeros of
$J_{l+1/2}\left(\epsilon_{n,l}\right)$ at fixed $l$, is the radial principal
quantum number. The degeneracy of each quantum mechanical energy level is
$g_{n,l}=2 l + 1$.

The counting function is defined as,
\begin{equation}\label{ns}
{\cal N}_{sph} (\epsilon) = g_s \sum_{n,l} g_{n,l} \ \Theta \left( \epsilon -
\epsilon_{n,l} \right) \ ,
\end{equation}
which increases by a factor $g_s \times g_{n,l}$ at each eigenvalue
$\epsilon_{n,l}$. The exact staircase function is represented in
Fig.~\ref{fig:3} ($g_s=2$), with the energy measured in units $\epsilon =
E/E_0$. Big jumps of the function at energy levels with large degeneracies are
clearly visible.

\begin{figure}
\centering
\includegraphics[height=4cm]{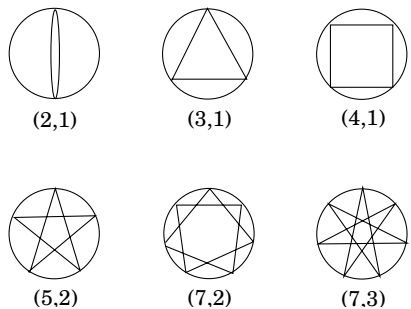}
\caption{Some of the periodic orbits of the sphere, labeled by $(v,w)$, the
number of vertices and the number of turns around the center, respectively.}
\label{fig:4}       
\end{figure}

We now turn to a semiclassical description of ${\cal N}_{sph} (\epsilon)$. A
classification and discussion of the periodic orbits of the sphere may be
found in \cite{bb}. Each orbit is characterized by two integers, $w$ (the
winding number), and $v$ (the number of vertices or bounces of an orbit off
the surface) (see Fig.~\ref{fig:4}). The semiclassical theory leads to an
approximate expression of the counting function in terms of these orbits,
given by Eqs.~(\ref{nss}), (\ref{nb}) and (\ref{no}). For the sphere, the
smooth and oscillatory parts read \cite{bb},
\begin{equation}\label{nssa}
{\overline {\cal N}}_{sph} (\epsilon) = \frac{2}{9\pi} \epsilon^{3/2} -
\frac{1}{4} \epsilon + \frac{2}{3\pi} \epsilon^{1/2} \ ,
\end{equation}
\begin{equation}\label{nsso}
{\widetilde {\cal N}}_{sph} (\epsilon) =
-2\sqrt{\epsilon}\ \sum_{w=1}^{\infty}\sum_{v=2w}^{\infty}
\frac{A_{vw}}{\ell_{vw}} \cos \left(\sqrt{\epsilon}\ \ell_{vw}+\nu_{vw}
\pi/2\right) \ .
\end{equation}
The different factors entering Eq.~(\ref{nsso}) are defined as,
\begin{equation} \label{para1}
A_{vw} = \left\{ \begin{array}{ll} - (2\pi w)^{-1} & \;\; v = 2 w \\ \\
(-1)^w \sin(2\ \theta_{vw})\sqrt{\sin(\theta_{vw})/\pi v} \ \epsilon^{1/4}
\;\;\;\;\;\; & \;\; v > 2 w \\
\end{array}
\right.
\end{equation}
\begin{equation} \label{para2}
\nu_{vw} = \left\{ \begin{array}{ll} 
0  & \;\;\;\; v = 2 w \\ \\
v+3/2
\;\;\;\;\;\;\;\;\;\;\;\;\;\;\;\;\;\;\;\;\;\;\;\;\;\;\;\;\;\;\;\;\;\;\;\;\;\;
\;\;\;\;\;\;\;\;\;\;\; & \;\;\;\; v > 2 w  \\
\end{array}
\right.
\end{equation}
\begin{equation} \label{para3}
\ell_{vw} = 2 v \cos(\theta_{vw}) 
\;\;\;\;\;\;\;\;\;\;\;\;\;\;\;\;\;\;\;\;\;\;\;\;\;\;\;\;\;\;\;\;\;\;\;\;\;\;
\;\;\;\;\;\;\;\;\;\;\;\;\;\;\;\;\;\;\;\;\;\;\;
\end{equation}
\begin{equation} \label{para4}
\theta_{vw} = \pi w/v \ .
\;\;\;\;\;\;\;\;\;\;\;\;\;\;\;\;\;\;\;\;\;\;\;\;\;\;\;\;\;\;\;\;\;\;\;\;\;\;
\;\;\;\;\;\;\;\;\;\;\;\;\;\;\;\;\;\;\;\;\;\;\;\;\;\;\;\;\;\;
\end{equation}
The parameter $\ell_{vw}$ is the length of the periodic orbit $(v,w)$ of a
unit radius sphere, and $\nu_{vw}$ is the phase attached to each orbit. The
semiclassical approximation of ${\cal N}_{sph}$ is also represented in
Fig.~\ref{fig:3} as a dotted line, together with the average part ${\overline
{\cal N}}_{sph}$. Viewed on large scales, the accuracy of Weyl's formula is
quite good. The inset shows a closer view of the function, where the
fluctuations with respect to the average and the accuracy of the semiclassical
approximation can be appreciated. Orbits with $w$ up to $40$ and $v$ up to
$80$ have been used in Eq.~(\ref{nsso}) to obtain the curve in the figure. It
is seen that the semiclassical approximation of ${\cal N}_{sph}$ is quite
good.

We now let for the moment the spherical cavity and return to general
considerations concerning the trace formula. Equation (\ref{rhoo}) shows that
the discrete quantum mechanical spectrum is, formally, recovered from an
interferent sum over all the periodic orbits whose contributions add up
constructively at the positions of the eigenvalues $E_j$, and destructively
elsewhere. Since in the vicinity of some reference energy $E_*$ the action may
be expressed as $S_p (E) \approx S_p (E_*) + \tau_p (E-E_*)$, each periodic
orbit contributes to the density with an oscillatory term, as a function of
$E$, whose wavelength is $h/\tau_p$. Long periodic orbits are therefore
responsible for the small scale structures of the spectrum. To resolve the
spectrum on an energy scale of the order $\delta$, and describe the departures
from the average properties on that scale, long periodic orbits of period
$\tauh$ are needed. In contrast, short orbits describe oscillations of the
density of states on large scales. The shortest orbit, of period $\taumin$,
defines the outer energy scale $E_c = h/\taumin$ of the density fluctuations.
The density modulations on a scale $E_c$ are usually referred to as shell
effects. Their importance heavily depends on the dynamical properties and
symmetries of the system. For instance, in systems with a central potential,
the angular momentum conservation produce strong degeneracies among
eigenvalues. The degeneracies create energy regions where the density of
eigenvalues is high compared to the mean, and others where it is small, thus
producing long range fluctuations of the density. This is an extreme
illustration of level bunching. From a classical point of view, the
enhancement produced by symmetries is due to the appearance of families of
periodic orbits all having, because of the symmetry, the same action,
amplitude, etc. This produces, at the level of the trace formula, an
enhancement of the shell effects. However, shell effects are present in
arbitrary systems, with or without symmetries, since they are a manifestation
of the existence of short periodic orbits, a generic property of any dynamical
system. In less symmetric systems they will be less important because the
degeneracy of the families of periodic orbits will be lower, or inexistent,
like in chaotic systems where periodic orbits are isolated. Correspondingly,
the associated shell effects will be less dramatic. This point will be
reconsidered in the next sections. See Ref.~\cite{sm}.

The scale $E_c$ is typically much larger than $\delta$. A precise
characterization of the difference is given by the adimensional parameter,
\begin{equation} \label{g}
g = E_c /\delta = \tauh/\taumin \ ,
\end{equation}
which counts the number of single--particle energy levels contained in a
shell. It is a measure of the collectivity of the long range spectral
modulations imposed by the shortest periodic orbits. It will play an important
role when quantifying the difference of size of shell effects in regular and
chaotic systems. Denoting $L = V^{1/3}$ the typical size of the system, it
follows from Weyl's law for a D--dimensional cavity that $g$ is proportional
to $(k L)^{D-1}$, where $k$ is the wavenumber at energy $E$. Thus, in the
semiclassical regime of short wavelength compared to the system size, $k L \gg
1$, the parameter $g$ is large.

To have a more precise idea of the consequences of the presence of an outer
scale in the oscillatory structure of the density, and of its relation with
the universal properties discussed in \S \ref{sec:2}, consider again the form
factor. Using the semiclassical approximation of the density,
Eq.~(\ref{rhoo}), it can be shown \cite{berry} that $K(\tau)$ is expressed as,
\begin{equation} \label{ffs}
K(\tau)= h^2 \left\langle \sum_{p,p'}\sum_{r,r'} A_{p,r} ~ A_{p',r'} \cos \left(
\frac{r S_p - r' S_{p'}}{\hbar} \right) ~ \delta \left[ \tau -
\frac{(r \tau_p+r' \tau_{p'})}{2} \right] \right\rangle _{E} \ ,
\end{equation}
where the brackets indicate an energy average (see \S \ref{sec:4.2} for a more
precise characterization of the averaging window).

For times $\tau \ll \tauh$ the off-diagonal contributions $p\neq p'$ in
Eq.~(\ref{ffs}) are eliminated by the averaging procedure, and the behavior of
$K(\tau)$ is well described by the diagonal terms $p = p'$. This gives a
series of delta peaks at $\tau = \tau_p$,
\begin{equation} \label{ffd}
K(\tau) \approx h^2  \sum_{p,r} A_{p,r}^2 ~ \delta \left( \tau - r \tau_p \right)  \ .
\end{equation}
The lowest peak is located at $\tau = \tau_{min}$, and for $\tau \leq
\tau_{min}$ the form factor is identically zero. Since the position and
amplitude of this peaks are system dependent, it follows that for times of the
order of $\tau_{min}$ no statistical universal behavior of the form factor
exists, and Eqs.~(\ref{krmtd}) and (\ref{kint}) do not provide a good
description. In particular, the fact that $K(\tau) = 0$ for $\tau \leq
\taumin$ is completely out of reach of the models based on random sequences.

\begin{figure}
\centering
\includegraphics[height=5cm]{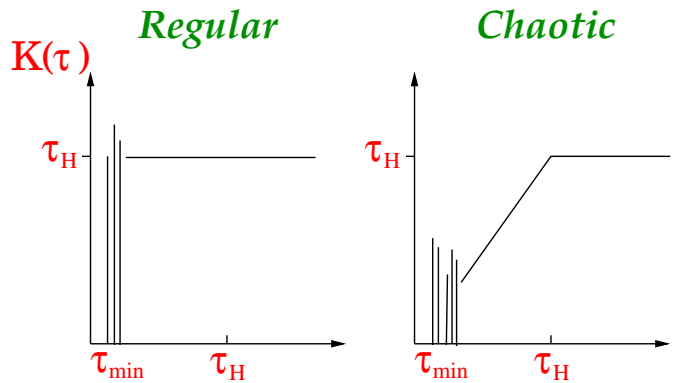}
\caption{Schematic view of the form factor for regular and chaotic systems}
\label{fig:5}       
\end{figure}

Going to longer times, but still remaining in the regime of validity of the
diagonal approximation, $\taumin \ll \tau \ll \tauh$, it can be shown that
indeed the RMT behavior (\ref{krmtd}) in chaotic systems, and the uncorrelated
statistics (\ref{kint}) in integrable systems, are recovered. Their
semiclassical origin is related to the statistical behavior of long classical
periodic orbits \cite{ho,berry}. There is no proof of the agreement for longer
times, aside some additional corrections in the $\tau$ expansion of the form
factor in time--reversal chaotic systems \cite{sieber}, and results in some
particular integrable models (e.g., the rectangular billiard, see
\cite{marklof}). We will therefore simply assume that for times $\tau \gg
\taumin$ the statistical description of the previous section, based on random
sequences, holds. Figure \ref{fig:5} summarizes the main features of the form
factor for integrable and chaotic systems.

In summary: the fluctuations of the single--particle spectrum in ballistic
systems are schematically of two types. On scales of order $\delta$, they are
universal (described by an uncorrelated sequence for integrable systems, and
by RMT in chaotic ones), as discussed in the previous section. In contrast, on
scales of order $E_c = g \delta$, with $g \gg 1$, there are long range
modulations whose structure and amplitude are system specific.

\section{Fermi Gas}
\label{sec:4}

We now turn to the central theme of these lectures, the theory of the quantum
fluctuations in interacting Fermionic systems. Our reference model will be a
confined Fermi gas, i.e. a gas of non--interacting particles moving in a
self--consistent mean field. It turns out, as it will be discussed below, that
such an approximation is well justified when studying the thermodynamic
fluctuations of interacting systems like
the atomic nucleus or the valence electrons in metals.

The starting point is a grand--canonical description, appropriate to study the
properties of an open system. Although this choice is in part motivated by
technical reasons, we will specify how to adapt the results to other physical
situations. In particular, to systems where the number of particles is
conserved, like the atomic nucleus.

When the chemical potential $\mu$ is fixed, the thermodynamic behavior of a
Fermi gas is described by the grand potential,
\begin{equation} \label{grand} \begin{array}{ll} 
\Omega (\mu,T,x) & = - \kb T \int dE ~\rho(E,x) ~ \log [1+e^{(\mu-E)/\kb T}]
 \\ \\
 & = - \int dE ~{\cal N} (E,x) ~ f(E,\mu,T) \ , \end{array}
\end{equation}
where $\rho (E,x)$ and ${\cal N} (E,x)$ are the single--particle density of
states and the counting function, respectively, already introduced in \S
\ref{2}. $T$ is the temperature, and $\kb$ is Boltzmann's constant. $x$
parametrizes the mean field potential, or represents some additional external
parameter, and $f$ is the Fermi function
\begin{equation} \label{f}
f(E,\mu,T) = \frac{1}{1+e^{(E-\mu)/\kb T}} \ .
\end{equation}

When the number of particles $A$ in the gas is fixed, the
energy of the system is
\begin{equation} \label{energy}
U (A,T,x)= \int dE~E~\rho(E,x)~f(E,\mu,T) \ , 
\end{equation}
where $\mu$ is determined from the equation
\begin{equation} \label{nmu}
A = \int dE~\rho(E,x)~f(E,\mu,T) \ . 
\end{equation}

Knowing these two thermodynamic potentials, other thermodynamic functions are
directly obtained by differentiation with respect to $\mu$, $x$, or $T$.

\subsection{Semiclassical approximation} \label{sec:4.1}

We now exploit the semiclassical expansion of the density of states to obtain
semiclassical expressions for the thermodynamic functions of the gas. Using
Eq.~(\ref{rhos}), the density of states in Eq.~(\ref{grand}) is written as a
sum of smooth plus oscillatory terms. This approximation is valid in the
semiclassical regime $A \gg 1$ where the typical wavelength of the fermions at
energy $\mu$ is much smaller than the system size, and is less accurate at the
bottom of the spectrum.

The substitution in Eqs.~(\ref{grand}) and (\ref{energy}) of the density of
states by its average behavior, Eq.~(\ref{rhob}) or (\ref{rhobc}), gives rise
to the smooth, bulk classical expressions for the thermodynamics of the Fermi
gas \cite{landau}. In finite fermionic systems, like the atomic nucleus or the
electron gas in a metallic particle, it is however well known that the smooth
part of the thermodynamic properties of the interacting system are not
correctly described by the Fermi gas approximation. In contrast, and this is a
crucial point, it is known that the {\sl fluctuations} of the properties of
the interacting system with respect to the average behavior and, in
particular, the energy fluctuations, are well approximated by the fluctuations
of the Fermi gas. This is the content of the Strutinsky energy theorem
\cite{strut}. This property justifies, in interacting many--body Fermionic
systems, the validity of the analysis of the fluctuations presented here,
based on a single--particle mean--field picture.

Equation (\ref{rhoo}) is then inserted in Eq.~(\ref{grand}) to compute the
oscillatory part of the grand potential. To leading order in $\hbar$ and for
low temperatures ($\kb T \ll \mu$, degenerate gas approximation) the integral
gives \cite{lm3},
\begin{equation}
\widetilde{\Omega} (\mu,T,x) = 2 \hbar^2 g_s \sum_p \sum_{r=1}^{\infty}
\frac{A_{p,r} ~ \kat(r ~ \tau_p)}{r^2 ~ \tau_p^2} \cos(r
S_p/\hbar+\nu_{p,r}) \label{grandosc} \ .
\end{equation}
In this expression, all the classical orbits have energy $E=\mu$, and they all
depend on the external parameter $x$. The temperature introduces the prefactor
$\kat(\tau)$,
\begin{equation}\label{kat}
\kat(\tau) = \frac{\tau / \tau_{\scriptscriptstyle T}}
{\sinh (\tau / \tauc)} \ , \;\;\;\;\;\;\;\;\;\;\;\;\; \tauc = h/ (2\pi^2 \kb T)
\ .
\end{equation}
This factor acts as an exponential cut off for orbits with period $\tau_p >
\tau_{\scriptscriptstyle T}$. For temperatures such that $\tau_{min} \gg
\tauc$, the quantum fluctuations are washed out and only the smooth part of
$\Omega$ given by ${\bar \rho}$ survives.

Similarly we may analyze the energy of the gas at a fixed number of particles.
The main technical difficulty here is that $\mu$ is not constant as some
external parameter is varied (like the shape of the nucleus for example), but
varies in order to satisfy, at any $x$, the condition (\ref{nmu}) ($\mu$ is in
fact a function of $A$, $T$, and $x$). It should therefore also be decomposed
into smooth and oscillatory parts,
\begin{equation}\label{mus}
\mu = \overline{\mu} + \widetilde{\mu} \ ,
\end{equation}
with $\overline{\mu}$ defined by the condition 
\begin{equation} \label{nmub}
A = \int dE~\overline{\rho}(E,x)~f(E,\overline{\mu},T) \ .
\end{equation} 
The fluctuating part of the energy of the system is defined as
\begin{equation}\label{ut}
\widetilde{U} = \int dE~E~\rho(E)~f(E,\mu,T) -
\int dE~E~\overline{\rho}(E)~f(E,\overline{\mu},T) \ .
\end{equation}
Using Eq.~(\ref{rhos}), assuming that $\widetilde{\mu} \ll \overline{\mu}$, and
integrating by parts Eq.~(\ref{ut}), one finds,
\begin{equation}\label{ut2}\begin{array}{ll}
\widetilde{U} = \int dE~E~\rho(E)~f'_{\bar{\mu}}
(E,\overline{\mu},T)~\widetilde{\mu} \ - & \int dE~\widetilde{\cal
N}(E)~f(E,\overline{\mu},T) \\ & - \int dE~E~\widetilde{\cal N}(E)~f'_E
(E,\overline{\mu},T) \ ,
\end{array}
\end{equation}
where $f'_{\bar{\mu}}(E,\overline{\mu},T) = d f(E,\overline{\mu},T)/d
\overline{\mu}$ and $f'_{E}(E,\overline{\mu},T) = d f(E,\overline{\mu},T)/d
E$. From Eqs.~(\ref{nmu}) and (\ref{nmub}), in a similar approximation, it
follows that
$$
\int dE~E~\rho(E)~f'_{\bar{\mu}}(E,\overline{\mu},T)~\widetilde{\mu} =
\int dE~E~\widetilde{\cal N}(E)~f'_E (E,\overline{\mu},T) \ .
$$
This implies,
\begin{equation}\label{utf}
\widetilde{U} (A,T,x) = - \int dE~\widetilde{\cal N}(E,x)~f
(E,\overline{\mu},T) \ .
\end{equation}
Comparing this equation with Eq.~(\ref{grand}), we conclude that the
fluctuations of the energy at a fixed number of particles $A$ coincide, to
leading order in an expansion in terms of $\widetilde{\mu}$, with the
fluctuations of $\Omega$ evaluated at a chemical potential $\overline{\mu}
(A,T,x)$ defined by inverting Eq.~(\ref{nmub}). From Eq.~(\ref{grandosc}), we
therefore get,
\begin{equation} \label{uosc}
\widetilde{U} (A,T,x) = 2 \hbar^2 g_s \sum_p \sum_{r=1}^{\infty}
\frac{A_{p,r} ~ \kat(r ~ \tau_p)}{r^2 ~ \tau_p^2} \cos(r S_p/\hbar+\nu_{p,r})
\ .
\end{equation}
All the classical orbits that enter this expression are at energy
$\overline{\mu} (A, T, x)$.

$\widetilde{U}$ are the variations of the energy due to kinematic effects
("dynamical fluctuations"). Equation (\ref{uosc}) is the main theoretical tool
to analyse its properties. It associates to each periodic orbit a certain
amount of energy that depends on its period, stability, etc. The total
contribution is the sum of all the energies associated to the periodic orbits.
The temperature enters only through the prefactor $\kat$, whose magnitude
decreases as $T$ increases. For studying the contribution of kinematic effects
to the mass of the nuclei, we simply take $T=0$. Then $\kat = 1$, and
Eq.~(\ref{uosc}) becomes,
\begin{equation} \label{uosc0}
\bt (A,x) = 2 \hbar^2 g_s \sum_p \sum_{r=1}^{\infty}
\frac{A_{p,r}}{r^2 ~ \tau_p^2} \cos(r S_p/\hbar+\nu_{p,r}) \ .
\end{equation}
This equation describes the fluctuations of the mass due to dynamical effects;
they are added on top of the smooth part given by the von Weizs\"acker
formula. At $T=0$, $\mu$ is the Fermi energy $\ef$ of the system, which is
also decomposed into a smooth and an oscillatory part, $\ef = \efa + \efo$.
The definition of the smooth part of the Fermi energy is, from
Eq.~(\ref{nmub}) at $T=0$, given by
\begin{equation} \label{nmubz0}
A = \int^{\ef} dE~\rho(E,x) = \int^{\efa} dE~\overline{\rho}(E,x) =
\overline{\cal N} (\efa,x) \ ,
\end{equation} 
which defines, by inversion, the function $\efa (A,x)$. All the classical
orbits in Eq.~(\ref{uosc0}) are evaluated at energy $\efa (A,x)$.

The oscillatory part of other thermodynamic functions may be computed by
direct differentiation of $\widetilde{\Omega}$ or $\widetilde{U}$ with respect
to the appropriate parameter \cite{lm3}.

\subsection{Statistical Analysis of the Fluctuations} \label{sec:4.2}

Equation (\ref{uosc0}) requires the knowledge of the periodic orbits and of
their different properties (period, stability, Maslov index, etc) in order to
compute the fluctuating part of the mass. In this Section, instead of a
detailed computation and description of the fluctuations for a particular
system, based on a precise knowledge of some set of periodic orbits, the aim is
to study their statistical properties. The interest of such an approach is
well known: using a minimum amount of information, a statistical analysis
allows to establish a classification scheme among the fluctuations of
different physical systems. It also allows to distinguish the generic from the
specific, and provides a powerful predictive tool in complex systems. A
general description of the statistical properties of the quantum fluctuations
of thermodynamic functions of integrable and chaotic ballistic Fermi gases,
and of their temperature dependence, was developed in Ref.~\cite{lm3}. Here we
will only concentrate on the energy of the gas at zero temperature, e.g., its
mass.

The fluctuating part $\bt (A,x)$ shows, as a function of the number of
particles or, alternatively, the Fermi energy $\efa$, oscillations described
by Eq.~(\ref{uosc0}). The statistical properties of $\bt$, and in particular
its probability distribution, are computed in a given interval of size $\Delta
\efa$ around $\efa$. This interval must satisfy two conditions. It must be
sufficiently small in order that all the classical properties of the system
remain almost constant. This is fulfilled if $\Delta \efa \ll \efa$. Moreover,
it must contain a sufficiently large number of oscillations to guarantee the
convergence of the statistics. As stated previously the largest scale
associated to the oscillations is $E_c$. Then clearly we must have $\Delta
\efa \gg E_c$. In the semiclassical regime the hierarchical ordering between
the different scales is therefore
$$
\delta \ll E_c \ll \Delta \efa \ll \efa \ .
$$

\subsubsection{Second Moment and Universality}

The average value of the fluctuating part $\bt$ defined as in
Eq.~(\ref{uosc0}) is zero. This is not strictly true in general, because it is
possible to show that subleading corrections in the expansion (\ref{utf}) have
a non--zero average. Ignoring this, the variance is the more basic aspect of
the probability distribution of the fluctuations. It provides the typical size
of the oscillations, and can easily be compared with experiments. We will now
compute a general expression for the second moment of the probability
distribution, that allows also to make an analysis of the universal properties
of the fluctuations. The key point is to understand which orbits give the
dominant contribution in Eq.~(\ref{uosc0}). Though it is clear that the weight
of an individual orbit decreases with its period, the net result of the sum of
the contributions is unclear, because the number of periodic orbits of a given
period grows with the period.

From Eq.~(\ref{uosc0}) the square of $\bt$ is expressed as a double sum over
the periodic orbits involving the product of two cosine. The latter product
may be expressed as one half the sum of the cosine of the sum and that of the
difference of the actions. The average over the term containing the sum of the
actions vanishes, due to its rapid oscillations on a scale $\Delta \efa$.
Therefore, letting aside for the moment the spin factor (which will be
discussed later on),
\begin{equation}
\langle \bt^2 \rangle = 2 ~ \hbar^4 \left\langle \sum_{p,p'}
\sum_{r,r'} \ \frac{A_{p,r}}{r^2 \tau_p^2} \ \frac{A_{p',r'}}{r'^2 \tau_{p'}^2} \
\cos \left( \frac{S_p - S_{p'}}{\hbar} \right) \right\rangle_{\efa,x} \ .
\label{vargen2}
\end{equation}
To simplify the notation we have included the Maslov indices in the definition
of the action. Ordering the orbits by their period, and taking into account
the restrictions imposed by the averaging procedure, the variance
Eq.~(\ref{vargen2}) can be related to the semiclassical definition of the form
factor $K(\tau)$, Eq.~(\ref{ffs}). The variance of the mass fluctuations takes
the simple form \cite{lm3,bl},
\begin{equation} \label{grandk}
\langle \bt^2 \rangle = \frac{{\hbar}^2}{2 \pi^2}
\int_0^{\infty} \frac{d\tau}{\tau^4} ~ K(\tau) \ .
\end{equation}
Analogous expressions connecting the variance of different thermodynamic
quantities, like for example the response of the gas to a perturbation, the
entropy, etc, at any temperature, can be found in Ref.~\cite{lm3}.

To obtain Eq.~(\ref{grandk}) we made use of the fact that the orbits giving a
non-zero contribution to (\ref{vargen2}) have similar actions (unless their
average will be zero). This implies that their period is also similar, and can
be considered to be the same in the prefactor (but not in the argument of the
oscillating part).

Based in Eq.~(\ref{grandk}) we now make a simple analysis of the variance of
the mass fluctuations. From Eqs.~(\ref{krmt}) and (\ref{krmtd}) it follows
that for chaotic systems the integrand in Eq.~(\ref{grandk}) behaves as
$\tau^{-3}$ for short times and $\tau^{-4}$ for long times, while
Eq.~(\ref{kint}) implies that for integrable motion the integrand varies as
$\tau^{-4}$. Therefore in both cases the integral (\ref{grandk}) converges for
long times. The dominant contributions come from short times, where the
integrand is large. If the form factor of a pure random sequence is used in
Eq.~(\ref{grandk}), the integral diverges. In real systems the divergence of
the integral is in fact stopped by the cutoff at $\tau = \tau_{min}$ of the
form factor. Because, as shown in \S \ref{sec:3}, the short-time structure of
the form factor is specific to each system (i.e., it is not universal), we see
from Eq.~(\ref{grandk}) that in general the second moment of the mass
fluctuations is non--universal, and consequently the same is true for the
probability distribution.

We thus conclude that {\sl the fluctuations of the mass are, regardless of the
regular or chaotic nature of the motion, dominated by the short non-universal
periodic orbits of the system}.

It should be mentioned that this is a particular property of the mass, that is
non generic, and therefore not shared by all other thermodynamic functions. It
can be shown that the fluctuations of some functions, like the entropy for
example, are dominated at low temperatures by times much larger than
$\taumin$; the corresponding distributions are universal. For a given
function, the universality can also depend on the nature of the dynamics. See
Ref.~\cite{lm3}.

\subsection{The second moment in the $\tau_{min}$--approximation} \label{sec:4.3}

We have demonstrated that the main contributions to the dynamical part of the
mass come from the shortest periodic orbits. On the other hand, we have seen
in \S \ref{sec:3} that for short times a good approximation to the form factor
(\ref{ffs}) is to keep only the diagonal terms in the double sum over the
periodic orbits, Eq.~(\ref{ffd}). Using the latter approximation, the variance
of the fluctuations of the mass is expressed as,
\begin{equation} \label{ud}
\langle \bt^2 \rangle \approx 2 ~ \hbar^4 
\sum_{p,r} ~ \frac{A_{p,r}^2}{r^4 \tau_p^4}  \ .
\end{equation}
This expression requires the explicit knowledge of the periodic orbits. This
information is not always available, like for example in the case of the
atomic nucleus. There is, however, a simple way to estimate the variance that
requires a minimum amount of information on the orbits. The approximation
consists in using in Eq.~(\ref{grandk}) the corresponding form factor (of an
uncorrelated sequence given by Eq.~(\ref{kint}) for a regular dynamics, of RMT
given by Eq.~(\ref{krmt}) for a chaotic motion), and to impose the additional
and important condition $K(\tau)=0$ for $\tau < \tau_{min}$ in all cases. This
is clearly an approximation, that we call the $\tau_{min}$-approximation. It
extrapolates the statistical behavior of the orbits down to times $\tau
\approx \tau_{min}$, ignoring the short-time system-dependent structures. All
the short-time structures are condensed into a single parameter, the period of
the shortest orbit.

The virtue of the $\tau_{min}$-approximation is to provide a simple estimate
of the size of the fluctuations, as well as of its dependence with the number
of particles, using a minimum amount of information.

Since the integral obtained from Eq.~(\ref{grandk}) in the
$\tau_{min}$-approximation is straightforward, we do not give here a detailed
account of its computation. The result, for integrable and chaotic systems,
given as an expansion in terms of the small parameter $1/g$, is,
\begin{equation} \label{ut3}
\langle \bt^2 \rangle = \left\{ \begin{array}{ll}
\frac{1}{24 \pi^4} g E_c^2 \;\;\;\;\;\; & {\rm Integrable} \\ \\
\frac{1}{8 \pi^4} E_c^2 \left(1-\frac{2}{g}+{\cal O} (g^{-2} \log g) \right) 
\;\;\;\;\;\; & {\rm Chaotic} \; \beta = 1 \\ \\
\frac{1}{16 \pi^4} E_c^2 \left( 1 - \frac{1}{3 g^2} \right)  & {\rm Chaotic}
\; \beta = 2 \ . \\
\end{array}
\right.
\end{equation}
It is evident from these expressions that the fundamental energy scale that
determines the energy fluctuations is $E_c$, and not $\delta$, or $\efa$. This
is natural, since we have shown that the energy or mass fluctuations are
controlled by the long--range fluctuations of the single--particle spectrum
produced by the short orbits on a scale $E_c$. Fluctuations of the levels in
smaller energy scales, whose statistical properties are universal as discussed
in \S \ref{sec:2}, contribute with $1/g$ corrections in Eq.~(\ref{ut3}). In
chaotic systems the variance of the fluctuations is twice smaller in systems
without time reversal symmetry. Semiclassically, this is also easy to
understand, because in systems with time reversal symmetry each orbit is
doubly degenerate (the primitive one and its time--reversed), in contrast to
systems with no time reversal symmetry. The coherent contribution of these
pairs of orbits produces a variance twice larger for $\beta=1$. Finally, the
variance of the fluctuations is $g$--times larger in integrable systems
compared to chaotic ones. This amplification, which could be quite large (a
precise estimate for nuclei is given in the next section), has as
semiclassical origin the existence of families of periodic orbits all
contributing in phase in Eq.~(\ref{uosc}). This estimate is valid for generic
integrable systems. Even larger amplifications could exist in
``super-integrable'' systems (cf next section).

Higher moments of the probability distribution of the energy fluctuations may
be computed similarly. Starting from Eq.~(\ref{uosc0}), a generalization of
the diagonal approximation used for the second moment can be implemented
\cite{lm3}. The results show that the moments of the distribution are
generically all different from zero, giving rise to asymmetric probability
distributions. For example, the third moment is given by,
\begin{equation} \label{m3b}
\langle \bt^3 \rangle \approx 6 ~ \hbar^6 \sum_p
\sum_{r_1=1}^{\infty} \sum_{r_2=1}^{\infty} {\cal A}_{p,r_1} ~ {\cal
A}_{p,r_2} ~ {\cal A}_{p,r_1+r_2} \cos
(\nu_{p,r_1}+\nu_{p,r_2}-\nu_{p,r_1+r_2}) \ ,
\end{equation}
where ${\cal A}_{p,r} = A_{p,r}/r^2 \tau_p^2$. This and the corresponding
expressions for higher moments were explicitly tested in some integrable and
chaotic models, were it is found that they work extremely well
\cite{lm3,lmb,lm4}.

\section{Nuclear Masses}
\label{sec:5}

The binding energy of a nucleus is defined by Eq.~(\ref{defb}). It is a direct
measure of the cohesion and stability of a nucleus. It is customary, as was
done in the previous sections, to analyze the experimental values observed by
decomposing ${\cal B}$ into two parts,
\begin{equation}\label{masc}
{\cal B} (E,x) = \ba (E,x) - \bt (E,x) \ .
\end{equation}
This splitting is at the basis of the so-called shell correction method,
introduced in nuclear physics by Strutinsky \cite{strut2} (see also
\cite{sm,bdjpsw}). Note the minus sign we have introduced in front of $\bt$.
The average part of ${\cal B}$ is described by a liquid drop model \`a la von
Weizs\"acker,
\begin{equation}\label{bv}
\ba = a_v A - a_s A^{2/3} - a_c \frac{Z^2}{A^{1/3}} -
a_A \frac{(N-Z)^2}{A} - a_p \frac{t_1}{A^{1/2}}  \ .
\end{equation}
The different terms in this expansion are associated to volume effects,
surface effects, Coulomb interaction, proton--neutron asymmetries, and pairing
effects, respectively. $N$ and $Z$ are the number of neutrons and protons,
$A=N+Z$ is the total nucleon number, and $t_1 = -1, 0, +1$ for even--even,
odd--even, and odd-odd nuclei, respectively. Note that all the terms are
smooth functions of $A$, $N$, and $Z$, except $t_1$ that takes into account
odd--even effects.

In contrast to $\ba$, the oscillating energy $\bt$, that describes deviations
with respect to the smooth part, can be analyzed, to leading order in a
one--body density expansion, in terms of the fluctuations computed from a
single--particle spectrum. This is the content of the Strutinsky energy
theorem \cite{strut}. As explained in \S \ref{sec:4}, this contribution is
related to kinematic aspects of the nucleus. It contains explicit information
about the motion of the nucleons in the self--consistent mean field. The net
output of the semiclassical theory for the oscillatory part of the binding
energy is Eq.~(\ref{uosc0}), that expresses $\bt$ as a sum of contributions,
each depending on a classical periodic orbit of the mean--field potential. Our
purpose now is to explicitly investigate the ability of such a formula to
describe the experimentally measured masses of the atomic nuclei.

\begin{figure}
\centering
\includegraphics[height=7.2cm]{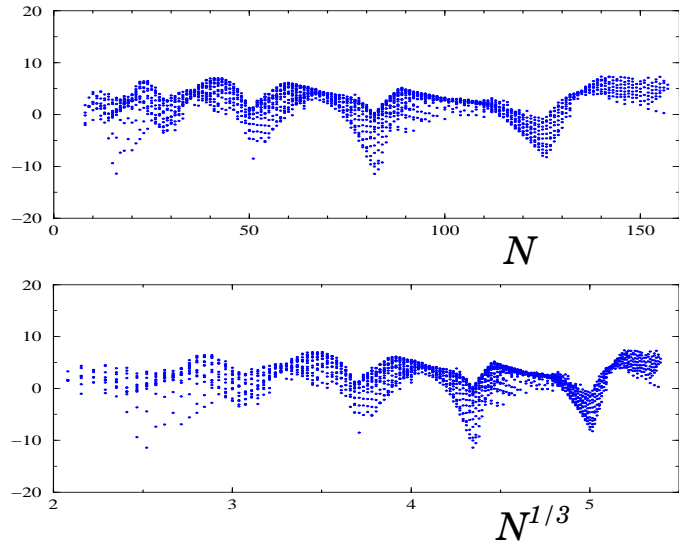}
\caption{The mass difference $\bt = \ba-{\cal B}$ (in MeV) for the 1888 nuclei
of the 1995 Audi--Wapstra compilation. Top: as a function of the neutron
number $N$. Bottom: as a function of $N^{1/3}$.}
\label{fig:6}       
\end{figure}

In order to proceed, the first thing to do is to subtract from the
experimental data the average part $\ba$. If this is done from the 1888 nuclei
with $Z,N\geq 8$ of the 1995 Audi--Wapstra compilation \cite{aw}, the result
is Fig.~\ref{fig:6}. The difference $\bt = \ba-{\cal B}$ is plotted as a
function of the neutron number (top part). Each point in the figure represents
a nucleus. The parametrization of $\ba$ used is taken from Ref.~\cite{prsz}:
$a_v=15.67, \ a_s=17.23, \ a_A=23.29, \ a_c=0.714$, and $a_p=11.2$ (all in
MeV).

A clear oscillatory structure (the shells) of amplitude $\sim 15$ MeV is
observed, with well defined sharp minima, that contrast with the more smooth
and rounded shape of the maxima. Another peculiar feature of the shells is
that the wavelength of the oscillations clearly increases with $N$. A minimum
in this plot corresponds to a minimum of $\bt$ (and therefore, locally, to a
maximum of ${\cal B}$). It defines more stable nuclei. The minima are located
at the "magic" numbers $N=28, 50, 82,$ and $126$ ($N=8$ and $20$ are not
clearly visible in this figure). The usual interpretation of these minima is
in terms of the existence of large gaps in the single--particle spectrum.
However, this definition is ill--defined, the notion of a large gap being
ambiguous. In the following, we will see that the most natural and simple
interpretation of all the characteristic features of this figure, including
the position of the magic numbers, is in terms of the periodic orbits of the
mean--field potential. This simple description does not pretend to be a
substitute of more elaborated many--body calculations. However, the
semiclassical theory clearly does something very important: it grasps the
essence of the physical mechanism responsible for the oscillatory structure of
the nuclear masses, and therefore provides a basic tool towards a general
understanding of shell effects.

\subsection{Regular motion: the spheroidal cavity}

To develop a semi--quantitative theory to explain the main features of
Fig.~\ref{fig:6}, a particular model needs to be specified. In the mean--field
approach adopted here, what is required is the shape of the mean--field
potential. Before selecting one, we need to make some general considerations
to guide our choices. In its ground state, the nucleus is minimizing its mass
(and therefore maximizing ${\cal B}$). The shell corrections are an important
component of the extremization procedure since, locally, they control it. The
type of dynamics the nucleons are undergoing depends on the form of the mean
field potential, through the parameter $x$. The function ${\cal B}(A,x)$ has
to be maximized with respect to $x$ at a fixed number of particles. Which type
of dynamics is energetically more favorable? As we have seen in \S
\ref{sec:4}, the amplitude of the shell effects is more important when the
motion of the nucleons is regular, by a factor $\sqrt{g}$, as compared to a
chaotic dynamics. To obtain a $\bt$ as large, and negative, as possible, a
mean field that produces a regular motion is therefore preferable.

For the mean--field approach and the semiclassical techniques employed here to
be meaningful, a large number of nucleons is required. In that case,
experiments show that a good approximation to the mean field is a potential of
the Wood-Saxon type, i.e., relatively flat in the interior of the nucleus and
steep on the surface \cite{bm}. To simplify, we will consider the potential to
be strictly flat in the interior (and set, by convention, to zero), and
infinite at the border. Our schematic model of the nucleus will therefore be a
mean--field potential given by a cavity, or billiard, with hard walls: the
nucleons freely move at the interior, and are elastically scattered off the
walls. The simplest geometry that produces a regular motion of the nucleons,
and that is of clear interest in nuclear physics, is a spherical cavity. We
will therefore start by considering in some detail the mass fluctuations of a
Fermi gas located at the interior of a 3D spherical cavity.

Quantum mechanically, at $T=0$ the fluctuating part of the energy of a Fermi
gas of $A$ particles in a spherical cavity of radius $R$ is given by,
\begin{equation}\label{bsph}
\bt_{sph} = \frac{\efa}{\epsf} \sum_{n,l}
\epsilon_{n,l} - \ba_{sph} \ .
\end{equation}
The normalized single--particle energies $\epsilon_{n,l}$ were defined in \S
\ref{sec:3}. The average part of the binding energy is given by $\ba_{sph}=g_s
(\efa/\epsf) [(2/15\pi) \epsf^{5/2} - (1/8) \epsf^2 + (2/9\pi) \epsf^{3/2}]$.
The sum in Eq.~(\ref{bsph}) is made over the lowest $A$ states of the sphere,
taking into account the spin ($g_s=2$) as well as the intrinsic ($
g_{n,l}=2l+1$) degeneracies. The energy factor in front of the sum is written,
for convenience, $\efa/\epsf=\hbar^2/2 m R^2 = E_0$, where
\begin{equation} \label{epsf}
\epsf = (\kfr)^2 = \efa/E_0 \ ,
\end{equation} 
$\efa = \hbar^2 \kfa^2 /2 m$ is the average Fermi energy, and $m$ is the nucleon
mass. Note that the normalized energy $\epsf$ is equivalent, up to a constant
factor, to the adimensional shell parameter $g$. Indeed, for a 3D cavity $g$ is
written, for $A \gg 1$,
$$
g=\frac{g_s}{\pi}\frac{V \kfa^2}{\ell_{min}} \ ,
$$
where $\ell_{min}$ is the length of the shortest periodic orbit. For a
sphere, with $V=(4/3) \pi R^3$ and $\ell_{min}=4 R$, this gives,
$$
g=\frac{g_s}{3} (\kfr)^2 = \frac{g_s}{3} \epsf \ .
$$

Semiclassically, the fluctuating part of the binding energy of the Fermi gas
is given by Eqs.~(\ref{uosc0}) and (\ref{nmubz0}). The former is obtained by
integrating with respect to the energy up to $\efa (A,x)$ the oscillating part
of the counting function. For a spherical cavity, $\widetilde{\cal N}$ is
given by Eq.~(\ref{nsso}). As usual, and to leading order in a semiclassical
expansion, only the integration with respect to the rapidly varying phase
factors is kept in the integral. As a result, the oscillatory part of the
binding energy of a Fermi gas in a 3D spherical cavity is,
\begin{equation}\label{bts}
\bt_{sph} (A,x) = 4 \ \efa g_s
\sum_{w=1}^{\infty}\sum_{v=2w}^{\infty} \frac{A_{vw}}{\ell_{vw}^2} \sin
\left(\kfr \ \ell_{vw}+\nu_{vw} \pi/2\right) \ .
\end{equation}
$A_{vw}$, $\ell_{vw}$, and $\nu_{vw}$ were defined in
Eqs.~(\ref{para1})--(\ref{para4}). In $A_{vw}$, the rescaled energy $\epsilon$
has to be replaced by the rescaled Fermi energy $\epsf$.

For simplicity, we consider separately the contributions to the binding energy
of the neutrons and of the protons. The contribution of the neutrons is given
by Eq.~(\ref{bts}) putting $A=N$. If the energy of the Fermi gas with a
spherical shape is computed as a function of the Fermi energy (grand-canonical
ensemble), it is found that the semiclassical approximation to the exact
result gives a very accurate description, of a quality comparable to what was
obtained in \S \ref{sec:3} for the counting function (cf Fig.~\ref{fig:3}).
The aim here, however, is to compute $\bt$ as a function of the neutron
number. The dependence of $\efa$ (or of $\epsf$) on $N$ is given, according to
Eq.~(\ref{nmubz0}), by the inversion of the smooth part of the counting
function. From Eqs.~(\ref{nss}) and (\ref{nssa}), the equation to invert is,
\begin{equation}\label{nsse}
\frac{N}{g_s} = \frac{2}{9\pi} \epsf^{3/2} - \frac{1}{4} \epsf +
\frac{2}{3\pi} \epsf^{1/2} \ .
\end{equation}

When the number of nucleons is large, only the first term in the r.h.s. of
Eq.~(\ref{nsse}) may be kept. Then,
\begin{equation} \label{kfr}
\kfr = \sqrt{\epsf} = \left( \frac{9 \pi}{2 g_s} N \right)^{1/3} = 1.92 \
N^{1/3} \ ,
\end{equation}
where we used $g_s=2$, the spin degeneracy. From Eq.~(\ref{epsf}), the
corresponding Fermi energy is,
\begin{equation} \label{ef}
\efa = \frac{\hbar^2}{2 m R^2} \epsf = 1.16 \frac{\hbar^2}{m r_0^2} =
40 \ \rm{MeV} \ ,
\end{equation} 
where we set the nuclear radius to $R = r_0 A^{1/3}$, $r_0 = 1.1$ fm, and
$m=939$ MeV. We have moreover assumed, for simplicity, an equal number of
protons and neutrons, $A= 2 N = 2 Z$.

Equation (\ref{kfr}) has a very simple and important consequence on the shell
oscillations. The oscillating part of the binding energy is given by a sum of
interferent terms each having, according to Eqs.~(\ref{bts}) and (\ref{kfr}),
a phase factor proportional to $N^{1/3}$. As a function of $N$, ${\widetilde
{\cal B}}_{sph}$ will therefore present oscillations whose wavelength grows as
$N^{2/3}$. Alternatively, a plot of the nuclear binding energy as a function
of $N^{1/3}$, instead of $N$, must show constant--wavelength oscillations.
This fact is confirmed, for the experimental data, in the bottom part of
Fig.~\ref{fig:6}.

To obtain a good semiclassical description of the shell effects it is not
enough to keep only the first term in the r.h.s. of Eq.~(\ref{nsse}). The full
equation should be used to compute $\epsf$ as a function of $N$ (this is
particularly important for the phase factors $\kfr$ in Eq.~(\ref{bts}), and
less for the prefactors). When this is done, a qualitative agreement between
the exact and the semiclassical results for the energy of a Fermi gas on a
sphere is obtained (see Fig.~12 in Ref.~\cite{mbc}). The precision is however
not satisfactory, and clear deviations between the exact and semiclassical
results are still observed. This is due to the intrinsic difficulty of
computing accurately the Fermi energy as a function of the number of
particles. The problem is particularly difficult (and probably one of the
worst cases that can be found) for the sphere, were large degeneracies are
present. These degeneracies make the function $\ef(N,x)$, defined by inverting
$N=\int^{\ef} \rho(E,x) dE$, ill defined.

There are different ways to cure this problem. A first possibility is to take
into account higher order terms in the expansion (\ref{utf}) in terms of
$\widetilde{\mu}$. The other method is to directly compute (numerically) the
integral in Eq.~(\ref{ut}), as was done in Ref.~\cite{mbc}. In this way, good
agreement between the quantum mechanical and the semiclassical results for the
sphere is obtained. This is the method employed here to compute
semiclassically the binding energies, shown in the figures below, as a
function of $N$. We will continue, however, to refer to the less accurate (but
qualitatively correct and more tractable) approximation (\ref{bts}) for
general considerations and discussions.

A priori, there are no adjustable parameters in the model. In particular, the
value of $\kfr$ is determined, for a given number of neutrons, by
Eq.~(\ref{nsse}). However, we find that a better agreement with the
experimental data is obtained if the value of the constant in Eq.~(\ref{kfr}),
$9\pi/2=14.1372$, is increased by ten percent, and changed to $15.625$. The
constant $9\pi/2$ is determined by the geometry of the sphere. The
modification can therefore be interpreted as a change that takes into account,
(i) the effects of a soft surface instead of a sharp border, and (ii) the
effect of spin--orbit scattering, that changes the lengths of the orbits in a
semiclassical description \cite{sos}. The modification of this constant is
equivalent to define an effective neutron number (cf Eq.~(\ref{kfr})),
$N_{eff}\approx 0.91 N$, and that is the way it was included in the
calculations.

\begin{figure}
\centering
\includegraphics[width=8.2cm,height=5.2cm]{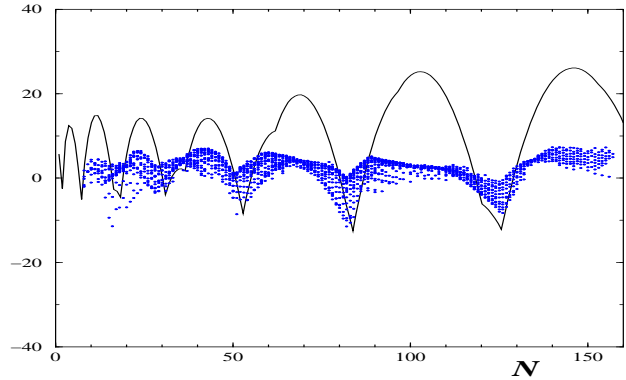}
\caption{The fluctuating part of the mass of the 1888 nuclei of the 1995
Audi--Wapstra compilation (dots) (cf Fig.~\ref{fig:6}), compared to the
fluctuating part of the energy of a Fermi gas on a spherical cavity (full
line).}
\label{fig:7}       
\end{figure}

We can now compare the fluctuating part of the energy obtained for a sphere
with the experimental data. The two quantities plotted are $\bt$ for the
experimental data (as defined by Eqs.(\ref{defb}) and (\ref{masc})), and
$\bt_{sph}$ for the theory. The result is shown in Fig.~\ref{fig:7}. It is
clear that the simple model of a spherical cavity (with only one adjusted
parameter ($\kfr$) that has been slightly increased) gives a pretty good
semi--quantitative description of all the main features of the shell effects
observed in the nuclear masses:
\begin{itemize}
\item[1)] the period of the oscillations, that increases with $N$, is very
well described by the model. The agreement between theory and experiment is
particularly good for $N\geq 25$, but clear correlations are also observed at
lower values of $N$.
\item[2)] the spherical cavity correctly reproduces the asymmetry between
sharp minima and rounded maxima.
\item[3)] the magic numbers obtained from the spherical model, at $N_{sph}=2,
7, 18$, $30, 52, 83, 126$, are in good agreement with the experimental values
$N_{exp}= 2, 8, 20,$ $28, 50, 82, 126$. As was shown in \S \ref{sec:4}, short
orbits dominate the fluctuations of the energy of the gas. In order to get an
analytic estimate of the magic numbers, we can simply consider the minima of
$\bt$ when only the triangular and square orbits are taken into account in
Eq.~(\ref{bts}). This gives the approximation,
\begin{equation} \label{magic}
N_{magic} \approx \left[ \frac{(2 k + 1) \pi}{10.77} \right]^3 
\approx 9, 18, 33, 55, 84, 122 \ldots
\end{equation}
$k$ is an arbitrary integer number, that has been taken equal to
$3,4,\ldots,8$ to obtain the values on the r.h.s. The constant $10.77$ is
computed from the parameters that define both orbits.
\item[4)] the amplitude of the oscillations is qualitatively correct. However,
in Fig.~\ref{fig:7} only the neutron contribution is plotted. Clearly, the
spherical cavity overestimates the amplitude.
\item[5)] The overestimate of the amplitude is larger for the maxima. This
difference is due to the well--known fact that the nucleus is deformed between
closed shells. This is treated in more detail below.
\end{itemize}

\begin{figure}
\centering
\includegraphics[width=10.5cm,height=8.0cm]{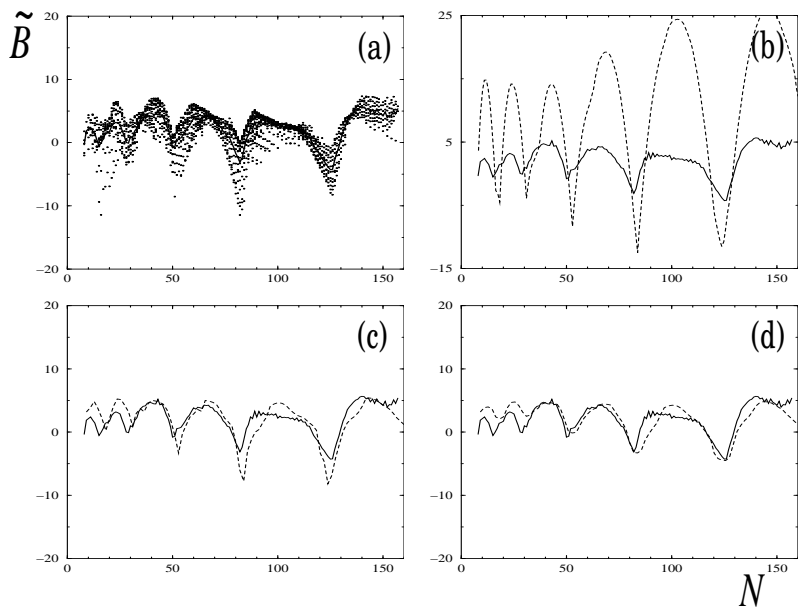}
\caption{Comparison between different theoretical approximations and
experimental values of the fluctuating part of the nuclear masses. (a)
Experimental data (dots), with their average behavior (full line). Average
experimental data (full line) compared to the predictions of (dashed line) (b)
a spherical cavity, (c) a (possibly) deformed cavity, and (d) a (possibly)
deformed cavity with a finite coherence length.}
\label{fig:8}       
\end{figure}

Figure \ref{fig:7} shows a direct comparison of the experimental data to the
binding energy of a Fermi gas on a spherical cavity as a function of the
neutron number. At each value of $N$, several experimental points appear in
the plot, that correspond to different atomic elements that vary by their
atomic number $Z$. In its present version, the model is too simple to
distinguish between different nuclei at a given $N$. The more sensible way to
proceed is to simply compare $\bt$ to the average of the experimental masses
computed at a fixed $N$. This average curve is shown, with respect to the
original experimental data, in part (a) of Fig.~\ref{fig:8}. In part (b) the
average curve obtained is compared to $\bt$ for a spherical cavity. This
figure is thus equivalent to Fig.~\ref{fig:7}, but with average experimental
results.

The negative peaks of $\bt$ define regions of more stable nuclei. In between
magic numbers, the spherical cavity leads to neutron numbers with a positive
contribution $\bt$. These large and positive contributions diminish the
binding energy. A possible mechanism to avoid that effect and to increase the
stability with respect to the spherical shape in between closed shells is {\bf
deformation}. By changing its shape, the nucleus can increase its stability by
finding regions were $\bt$ is more favorable. To complete the previous model,
it is therefore necessary to incorporate symmetry breaking degrees of freedom.
In the new space of deformations, the mass is determined by minimizing, at a
fixed particle number, the binding energy with respect to the deformation
parameters.

The deformation is here treated as a small perturbation of the spherical shape
\cite{creagh}. An important technical advantage of the perturbative approach
is that it leads to a very simple semiclassical treatment, where the
deformation is incorporated as a modulation factor into the previous formulas:
$A_{p,r}$ is replaced by $A_{p,r} \times M_{p,r} (x)$ in Eq.~(\ref{bts})
\cite{creagh}. $x$ denotes the deformation parameters, taken as quadrupole
($x_2$), octupole ($x_3$), and hexadecapole ($x_4$) axial symmetric
deformations of the surface of the sphere parametrized, in spherical polar
coordinates $(r,\theta,\phi)$, by $r(\theta) = R \left[ 1 + x_0 +
\sum_{j=2}^{4} x_j P_j (\cos \theta)\right]$. $x_0$ is included to ensure
volume conservation, and the $P_j (y)$ are the Legendre polynomials of order
$j$. The explicit expressions of the modulation factors $M_{p,r} (x)$ for the
three deformations of the periodic orbit families of a spherical cavity may be
found in Ref.~\cite{mbc}. Including these factors in the semiclassical
expressions, at a fixed neutron number $N$, the minimum of the binding energy
is located in the parameter space $(x_2,x_3,x_4)$.

The results of the minimization procedure for nuclei with neutron number up to
160 are shown in part (c) of Fig.~\ref{fig:8}. With respect to a perfectly
spherical shape, the results are greatly improved and the agreement with the
experimental data is much better. The amplitude of the shell corrections have
been strongly diminished. The shape and position of the negative peaks of
$\bt$ remain largely unchanged. The main difference is that the large positive
peaks have been suppressed, and replaced by a less pronounced oscillatory
behavior.

As a final improvement of the model we consider {\bf inelastic processes}. The
model presented above is based on a mean field approach, were each nucleon (or
quasiparticle) freely moves in the self--consistent mean field produced by the
other nucleons. This single--particle picture is of course an idealized
approximation of the full many body problem. In reality, quasiparticles have a
finite lifetime associated to inelastic processes. By producing
single--particle energy levels with a certain width, nucleon--nucleon
inelastic scattering tends to wash out the coherent phenomena that lies behind
the shell effects, and manifests as a phase--breaking mechanism. As a
consequence, a finite phase coherence length enters now the play, that
corresponds to the typical distance a quasiparticle travels without loosing
phase coherence.

When the finite width of the single--particle levels is included into the
semiclassical treatment, the net output is very natural. It manifests as a
damping factor in the sum over periodic orbits, that suppresses the
contribution of long periodic orbits \cite{bb}. This is similar to what
happens to shell effects as the temperature of the system is raised
\cite{lm3}. In the approximation (\ref{bts}), the fluctuating part of the
energy, including deformations as well as inelastic processes, is obtained by
replacing $A_{p,r}$ by $A_{p,r} \times M_{p,r} (x) \times
\kappa_{\xi}(\ell_{vw})$, where $\kappa_{\xi}(\ell) = (\ell/\xi)/\sinh
(\ell/\xi)$, and $\xi$ is the phase coherent length expressed in units of the
radius of the sphere. For orbits whose length $\ell_{vw} \gg \xi$, the latter
factor produces an exponential damping of the corresponding contribution. In
the opposite limit, when $\ell_{vw} \ll \xi$, $\kappa_{\xi}$ tends to one. The
phase coherence length $\xi$ measures the typical number of times a
quasiparticle can bounce back and forth in the mean field potential before
loosing phase coherence.

Fixing this length to $\xi \sim 5$, the fluctuating part of the binding energy
$\bt$ is represented in part (d) of Fig.~\ref{fig:8}. With respect to part
(c), which does not include decoherence phenomena, the amplitude and peaks are
diminished, and now the theoretical description provides a very convenient
description of the average experimental data. The agreement is less good at
small neutron numbers $N \leqa 30$, a foreseeable discrepancy from a theory
based on a large $N$ expansion. The RMS error between the theoretical curve of
Fig.~\ref{fig:8}(d) and the average experimental data is of $1.3$ MeV (taking
into account only points with $N \geqa 30$). This error is only a factor two
larger than the best current results of global mass adjustments, a remarkable
result considering the simplicity of the model.

Before closing this section on the ability of a spheroidal cavity to reproduce
the experimental data, we would like to consider, within the present model,
{\bf supershell structures}. These are long--range coherent modulations of the
amplitude of the shell oscillations. They constitute a further remarkable
manifestation of the collective deviations of the single--particle spectrum
with respect to its average behavior. Initially predicted for a spherical
cavity by Balian and Bloch \cite{bb,nhm}, they were observed experimentally in
metallic clusters \cite{knight}. In nuclear physics, in principle the range of
variability of the number of nucleons is too small in order to clearly display
the effect. However, we will see that, contrary to this expectation, there
exist some indications of this effect in the nuclear masses.

\begin{figure}
\centering
\includegraphics[width=6.5cm,height=5.5cm]{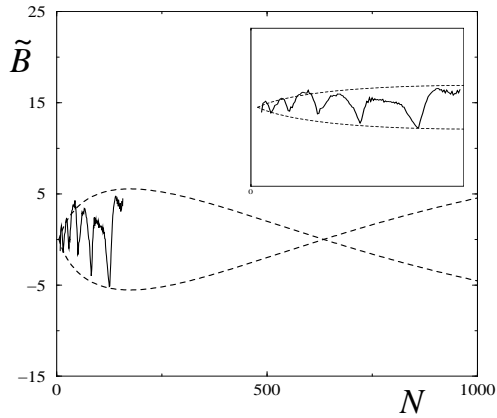}
\caption{The average fluctuating part of the mass of the 1888 nuclei of the
1995 Audi--Wapstra compilation (full line) compared to the modulation factor
of the shell oscillations (supershell envelope, dashed line). The inset shows
a closer view of the same curves, where the node at $N\approx 5$ is visible.}
\label{fig:9}       
\end{figure}

The simplest and more elegant formulation of supershells is based on
semiclassical arguments. They are associated to the beating pattern produced
by the interference of two periodic orbits of the spherical cavity that have
similar lengths, the triangle (3,1) and the square (4,1) \cite{bb,nhm}. When
only their contributions are taken into account in Eq.~(\ref{bts}), the
oscillatory part of the binding energy is approximated by
\begin{equation}\label{ssh}
\bt \approx C N^{1/6} \cos \left( \frac{\varphi_{\scriptscriptstyle
\triangle}-\varphi_{\scriptstyle \diamond}}{2} \right)
\cos \left( \frac{\varphi_{\scriptscriptstyle \triangle} +
\varphi_{\scriptstyle \diamond}}{2} + \frac{\pi}{2}
\right) \ ,
\end{equation} 
where $C$ is some amplitude factor that has no relevance in the present
discussion, $\varphi_{\scriptscriptstyle \triangle}=3\sqrt{3} \times 1.98
N^{1/3} + 9\pi/4$ and $\varphi_{\scriptstyle \diamond}=4\sqrt{2} \times 1.98
N^{1/3} + 11\pi/4$ are the phases $\kfr \ \ell_{vw}+\nu_{vw} \pi/2$ associated
to the triangle and the square orbits, respectively. This simple equation
describes the supershell effect. On the one hand it includes the shells, given
by the fast oscillations produced by the second cosine in Eq.~(\ref{ssh}) (sum
of the phases). The approximate magic numbers Eq.~(\ref{magic}) were computed
by locating the minima of this term. The amplitude of the fast oscillations is
modulated, on long range scales, by the term containing the difference of the
phases. These modulations produce a beating pattern of the shell amplitudes.
The nodes of this pattern determine the particle numbers were the shells have
a minimum amplitude. From Eq.~(\ref{ssh}), the nodes are located at
\begin{equation} \label{sshn}
N_{nodes} \approx \left[ \frac{(2 n + 1/2) \pi}{(4\sqrt{2}-3\sqrt{3})\ 1.984}
\right]^3 \approx 5, 634, 3699, \ldots
\end{equation} 
where $n=0,1,2,\ldots$ is an arbitrary integer number. Atomic nuclei of 630
neutrons are thus necessary to observe one supershell oscillation. In spite of
this difficulty, we can however explore if there are traces of the modulation
in the available experimental data, having in mind that a node is predicted
for $N\approx 5$. For that purpose, we plot in Fig.~\ref{fig:9} the average
experimental curve (defined in part (a) of Fig.~\ref{fig:8}), and compare it
to the modulation factor $C N^{1/6} \cos \left[ \left(
\varphi_{\scriptscriptstyle \triangle} - \varphi_{\scriptscriptstyle \diamond}
\right)/2 \right]$.

Although the experimental data is indeed not unambiguous due to the limited
range of the number of neutrons, it is clear that they follow, to a good
approximation, the supershell modulations. It should be noted that the
increase of the amplitude observed when going from $N\sim 10$ to $N\sim 160$
(better displayed in the inset) cannot be explained by the overall factor
$N^{1/6}$ in Eq.~(\ref{ssh}). It requires, as a main ingredient, the
supershell modulation factor.

The perturbative treatment of a spherical cavity, with an underlying classical
regular motion of the nucleons, allows therefore to understand in very simple
physical terms most features of the shell fluctuations observed in the atomic
masses. Beyond this specific model, it is possible to use the general
arguments introduced in the previous section to estimate the typical size of
the shell effects in the nuclear masses to be expected from a generic
integrable motion. In the experimental curve, Fig.~\ref{fig:6}, the
fluctuations have approximately constant amplitude, with an RMS size
$\sigma_{exp} \approx 3$ MeV.

A theoretical prediction of the typical size of the fluctuations can be made
from the results of \S \ref{sec:4} (cf Eq.~(\ref{ut3}) and Ref.~\cite{bl}).
The variance of the binding energy or mass fluctuations of a Fermi gas whose
corresponding classical dynamics is regular is given by,
\begin{equation} \label{utreg}
\langle \bt_{reg}^2 \rangle = \frac{1}{24 \pi^4} g E_c^2 \ .
\end{equation}
We thus simply need to compute $g$ and $E_c$ for the nucleus. The length of
the shortest periodic orbit can be estimated as twice the diameter of the
nucleus. Its period is $\taumin = 4 R/v_{\scriptscriptstyle F}$, where $R$ is
the nuclear radius and $v_{\scriptscriptstyle F}$ the Fermi velocity. Then,
according to the definition (\ref{ec}),
$$
E_c = \frac{h}{\taumin} = \frac{h v_{\scriptscriptstyle F}}{4 R} = 
\frac{\pi \efa}{\kfr} \ .
$$
From Eq.~(\ref{kfr}) (or its generalization to an arbitrary shape), with
$N=A/2$, we have $\kfr \approx 1.5 A^{1/3}$. Therefore, using moreover
Eq.~(\ref{ef}), we obtain,
\begin{equation}\label{ecreg}
E_c = \frac{80}{A^{1/3}} \ \rm{MeV} \ .
\end{equation} 
On the other hand, from Eqs.~(\ref{rhobc}) and (\ref{nb}), putting
$\overline{\cal N}=A$, it follows that
$$
\delta = \frac{2}{3} \frac{\efa}{A} \ ,
$$
and hence,
\begin{equation}\label{greg}
g = \frac{E_c}{\delta} \approx 3 \times A^{2/3} \ .
\end{equation}
In Eq.~(\ref{utreg}), the $A$ dependence between $E_c^2$ and $g$ cancels
exactly, and the RMS of the mass fluctuations of a regular motion is, finally,
\begin{equation} \label{varreg}
\sigma_{reg} = \sqrt{\langle \bt_{reg}^2 \rangle} \approx 
2.9 \ \rm{MeV} \ ,
\end{equation}
which is in good agreement with the experimental result. For a generic regular
dynamics, the size of the mass fluctuations is therefore expected to be
constant, e.g. with no dependence on the number of particles. In this respect,
the spherical cavity is special. In fact, Eq.~(\ref{para1}) shows that the
amplitude of all the periodic orbits (except the $v=2 w$ family, which is of
lower weight) scales as $\epsf^{1/4} \sim A^{1/6}$ (cf Eq.~(\ref{kfr})).
Therefore, the size of the fluctuations for a spherical cavity scales as
$\sigma_{sph} \sim A^{1/6}$. In atomic nuclei, this prediction is not
generically valid, due to deformations. But this scaling is expected to hold
for the strength of the peaks associated to the magic numbers, where the
nucleus is well described by a spherical shape. From a dynamical point of
view, the specificity of the sphere compared to a generic regular motion comes
from the fact that the codimension of the $v>2 w$ families of periodic orbits
of the sphere is three, whereas the generic codimension for 3D integrable
systems is two.

\subsection{Chaotic motion}

\begin{figure}
\centering
\includegraphics[height=6.2cm]{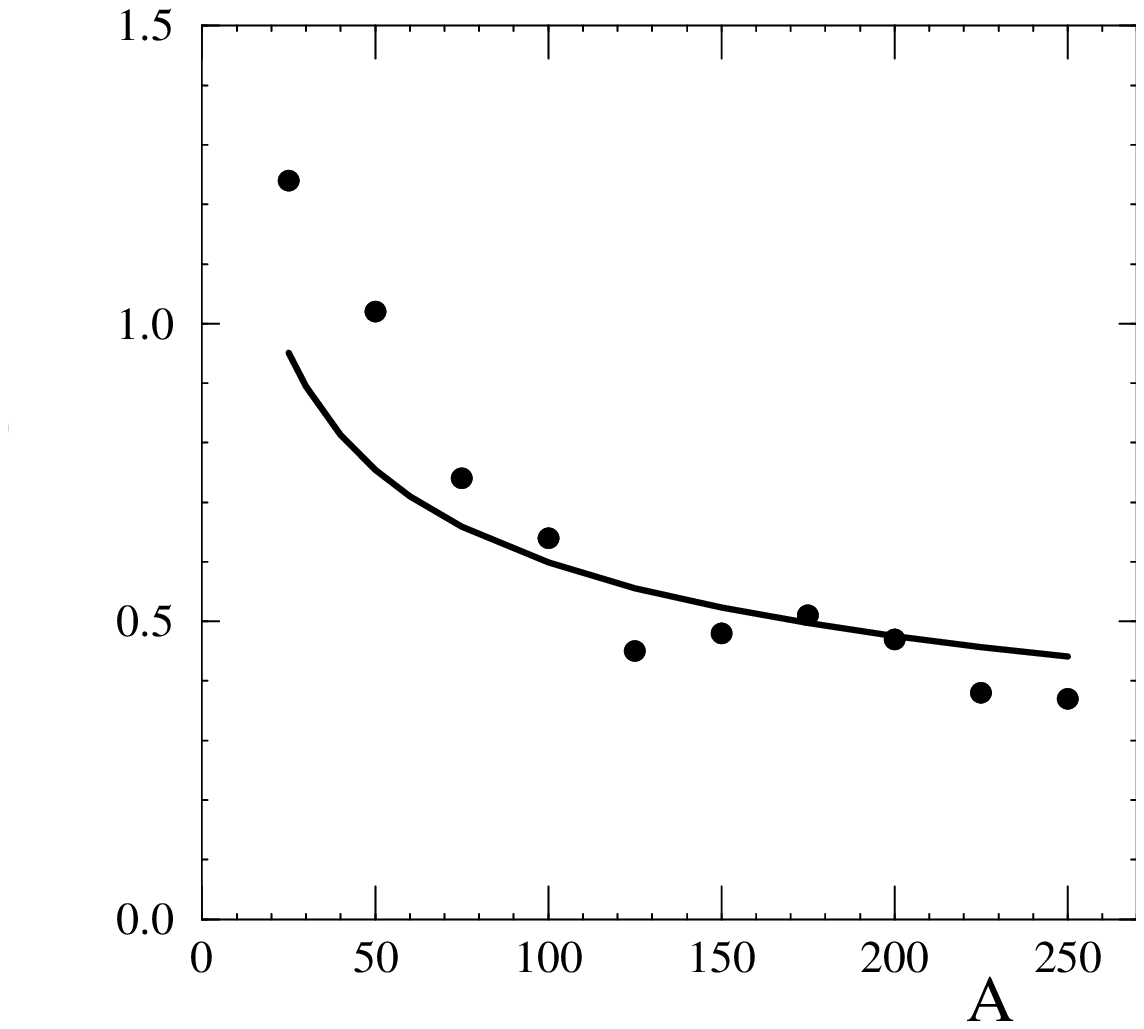}
\caption{The RMS of the difference between computed and observed masses as a
function of the mass number (dots, from Ref.~\cite{mnms}), compared to the RMS
of the mass fluctuations due to a chaotic motion. Taken from Ref.~\cite{bl}.}
\label{fig:10}       
\end{figure}

The description of the mass fluctuations in terms of a regular nucleonic
motion is globally quite satisfactory. However, the semiclassical methods
employed here only provide a semi--quantitative description. Actual
calculations of the masses are in fact much more sophisticated. Different
methods, more or less phenomenological, have been used to produce global
systematics of the nuclear masses. These are based on shell model approaches,
mean field theories (with or without correlations, using different effective
forces), or phenomenological liquid drop models plus shell corrections. The
number of adjustable parameters in these theories is considerable, of the
order of 20-30. See Ref.~\cite{lpt} for a recent review that describes the
different approaches. In spite of the sometimes very different conceptual
basis of these models, there is a peculiar common feature that emerges. Over
the whole mass chart of known nuclides, the different models possess a
comparable precision, of the order of $400-700$ keV. A typical RMS error with
respect to the experimental data, taken from Ref.~\cite{mnms}, is shown in
Fig.~\ref{fig:10} (dots). An error of the order of 500 keV, which decreases
with $A$, is observed.

Is there a way to understand this deviation? The fact that different models have
approximately the same precision points to the fact that there may be a
fundamental physical mechanism at work here. In the present framework based on
a semiclassical description of the motion of the nucleons, a natural scheme to
explore is the presence of some chaotic motion. Although we have argued that a
regular motion of the nucleons is energetically more favorable than a chaotic
one, there are several reasons that make the presence of the latter
inevitable. The first one is that a perfectly regular motion is rather unusual
in nature (it holds for the two--body problem with central forces, but the
three body is already not integrable). A mixed dynamics, where regular and
chaotic motion coexist, is generically expected to occur. There is no simple
reason to believe that the full many--body nuclear problem, with of the order
of $100$ interacting particles, will not follow the generic rule. The second
argument in favor of the presence of chaos is also fundamental. As we will
argue below, there are indications that what is really computed by assuming
some chaoticity in the nucleon's motion are correlation effects that are
beyond the mean--field description of the nucleus.

We therefore assume that the phase--space of the nucleons at Fermi energy is
dominated by regular components but contains, nevertheless, some chaotic
layers. The idea is to compute the contribution to the mass of these chaotic
layers, and then compare it to the deviation observed between the experimental
data and the best theoretical predictions. From a semiclassical point of view,
the most simple approximation that can be made in the generic case of a mixed
dynamics is to split the sum over periodic orbits in Eq.~(\ref{uosc0}) into
two terms, one from the regular part, the other from the chaotic phase space
sector. The mass fluctuations are now written as,
\begin{equation} \label{btrc}
\bt = \bt_{reg} + \bt_{ch} \ .
\end{equation}
The two terms on the r.h.s. are, from a statistical point of view,
independent, $\langle \bt_{reg} \bt_{ch} \rangle =0$. This happens because, as
already pointed out in \S \ref{sec:4}, the dominant contribution to the energy
comes from the short orbits, with $\tau_p \ll \tauh$. Since the orbits
contributing to each term are different because they lie in different
phase--space regions, the cross product of two different orbits vanishes by
the averaging procedure (assuming that the actions of the orbits are
incommensurable).

We now need to evaluate the variance of the shell corrections that originate
in the chaotic layers of a nucleus, given by the term $\bt_{ch}$ in
Eq.~(\ref{btrc}). We want to make a statistical estimate which, as discussed
before, has the advantage of requiring a minimum amount of information about
the system. This can be done from the general result obtain in \S \ref{sec:4},
Eq.~(\ref{ut3}) with $\beta=1$ (assuming time reversal symmetry),
\begin{equation} \label{utch}
\langle \bt_{ch}^2 \rangle = \frac{1}{8 \pi^4} E_c^2 \ .
\end{equation}
An estimate of $E_c$ is now required. In fact, the result obtained previously,
Eq.~(\ref{ecreg}), can be used here. The reason is that $E_c$ is defined as
the energy conjugate to the shortest periodic orbit. In a system were regular
motion coexists with chaos, the shortest orbit lying in the regular or in the
chaotic components will have approximately the same period (or length) at
fixed energy $\efa (A,x)$. Therefore, the scale $E_c$ associated to each of
the contributions in Eq.~(\ref{btrc}) will be similar.

No other information is needed to estimate the size of the contributions to
the mass that originate from the chaotic layers. Using Eq.~(\ref{ecreg}), we
obtain,
\begin{equation} \label{varch}
\sigma_{ch} = \sqrt{\langle \bt_{ch}^2 \rangle} \approx \frac{2.8}{A^{1/3}} \
{\rm MeV} \ .
\end{equation}
This result has to be compared with the RMS of the difference between the
experimental and the computed masses, which we take, as mentioned before, from
Ref.~\cite{mnms}. The comparison is shown in Fig.~\ref{fig:10}.

The agreement is quite good. The amplitude in Eq.~(\ref{varch}) is uncertain
up to an overall factor of say, $2$. It can be varied by increasing slightly
the period of the shortest orbit (we have chosen twice the diameter of the
nucleus, which is the shortest possible length; any modification, assuming for
example a triangular--like shape, or any other shape, will increase the length
and period, and therefore will diminish $\sigma_{ch}$), and by a more
appropriate inclusion of spin and isospin (this increases $\sigma_{ch}$ by a
factor $2$ if these components are treated as uncorrelated). The $A$
dependence of the error is very well fitted in the region $A \geqa 75$, with
deviations observed at lower mass numbers. This is consistent with the
expected loss of accuracy of the semiclassical theories for light nuclei.

There are several features of the present theory -- that interprets the mass
deviations in terms of chaotic motion -- that make its predictions reliable.
First of all, Eq.~(\ref{utch}) contains only {\sl one physical parameter}, the
period of the shortest chaotic periodic orbit, which is a function of $A$
because the size of the nucleus increases with the mass number. It has no
dependence on the relative size of the chaotic region, i.e., on the fraction
of phase space occupied by chaotic motion. Without this quite remarkable and
important fact we would have been forced to estimate that fraction, something
that is hardly possible with present knowledge, despite the efforts in this
direction since the pioneering work of Ref.~\cite{arvieu}. Mathematically,
this is due to the fact that, in chaotic systems, the dominant part of the
form factor in the integral (\ref{grandk}) is independent of the Heisenberg
time (cf Eq.~(\ref{krmtd})). But it is precisely that time that has
information on the corresponding level density $\delta=\rhob^{-1}$ (the total
average level density of a mixed system is the sum of the average level
densities of the regular and chaotic components. Each of these densities is
proportional to the phase space volume occupied by the corresponding motion).
Therefore, for the chaotic component the integral (\ref{grandk}) has no
information on the relative phase space volume occupied by the chaotic layers.
This information enters only as a correction, through the parameter $g$ in
Eq.~(\ref{ut3}).

Equation (\ref{utch}) is in fact quite robust. It is not only independent of
the chaotic phase--space volume, but it is also valid for arbitrary
dimensions. After a system--dependent transient (schematically represented in
Fig.~\ref{5}), the linear growth of the form factor in Eq.~(\ref{grandk}) for
chaotic systems with $\beta=1$ is a quite general feature, independent of {\sl
any} information of the system. This fact suggests a very natural origin of
these fluctuations. Although our analysis is based on a single--particle
picture, it can be extended to the full many--body phase space (of dimension
$6 {\scriptstyle \times} A$). In that space, and in a semiclassical picture,
to first approximation the point representing the system follows a very simple
trajectory, driven by the regular mean--field that dominates the motion. On
top of that, it is likely that the residual interactions, not taken into
account in that approximation, induce chaotic motion. The presence of chaotic
orbits would then introduce additional long--range modulations.
Eq.~(\ref{utch}) would then still be valid to evaluate the amplitude of those
modulations, with $\tau_{min}$ the period of the shortest chaotic orbit in the
multidimensional space. Rough estimates indicate that $\tau_{min}$ is
comparable to the three--dimensional non--interacting period (this
corresponds, for example, to collective and coherent oscillations of the
particles). Therefore Eq.~(\ref{utch}) presumably gives an estimate of the
mass fluctuations arising from neglected many--body effects. This deserves
further investigation, though some work in this direction has already been
undertaken \cite{fw}.

\section{Conclusions}
\label{sec:6}

We have presented in these lectures a study of shell effects on nuclear ground
states within a unified framework, namely periodic orbit theory. In a
mean--field approximation of the nuclear dynamics, the single--particle motion
is quantized. The regular or chaotic nature of the classical single--particle
motion imprints the single--particle spectrum in two different ways. The first
one occurs at the scale of the mean level spacing $\delta$, and produces
different (but universal) statistical fluctuations. The second acts on a much
larger scale $E_c$, set by the inverse of the time of flight across the
system. A bunching of the single--particle levels produced by the short
periodic orbits is observed on this scale, whose intensity depends on the
regular or chaotic nature of the motion. These two features of the
single--particle spectrum influence the probability distribution of the
quantum fluctuations of thermodynamic functions of the Fermi gas in different
ways. In the particular case of the mass, the fluctuations are controlled by
the short periodic orbits.

Without need of heavy and numerically expensive many--body calculations, the
(phenomenological) semiclassical theory, although less precise, provides a
good semi--quantitative explanation of most of the features of the
experimental data on nuclear masses. A simple model of regular classical
motion based on a spherical cavity (including the possibility of deformations
and decoherence effects) gives already a good agreement. Within this model,
the average behavior of the experimental data is reproduced, for nuclei with
more than 30 neutrons, with a RMS error of $1.3$ MeV. This error is only a
factor two larger than the best current results of global mass adjustments, a
remarkable result considering the simplicity of the model. This stresses that
the main physical effect behind shell effects is purely kinematic, with
spin--orbit or other interactions having a small overall influence. The latter
were phenomenologically taken into account by adjusting the value of $\kfr$.
Within this simplified model, long range modulations of the shell amplitudes
(supershells) were also investigated. Although, due to a limited particle
number, this effects cannot unambiguously be displayed, clearly the present
experimental data show some indications of supershell structure in the atomic
masses.

When the experimental data is compared with the best global theoretical
calculations over the whole mass spectrum, deviations of the order of $500$
keV, which decrease with $A$, are observed. We interpret this deviation as due
to the presence of some chaotic components in the phase space dynamics of the
nucleons. Our computations require very little information on the system. In
spite of that, they are in good agreement with the deviations observed. Our
interpretation does not imply the existence of a ``chaotic'' insurmountable
obstacle that, intrinsically, limitates our understanding and predictive power
to compute the nuclear masses (see, in this respect, Ref.~\cite{aberg}). It
simply says that at these scales (of the order of $500$ keV) a new effect
appears that needs to be taken into account properly. The picture we are
suggesting is coexistence of order and chaos as produced, for instance, by
residual interactions. Further evidence should be given to confirm this
scenario. In particular, the theory predicts autocorrelations in the total
energies \cite{lm3}, as well as the effect of the presence of chaotic layers
on the nuclear level density (as a function of the excitation energy).


%
%
%

\begin{thebibliography}{99.}
\bibitem{foot1} If the system has discrete symmetries (like the parity
symmetry in the cardioid billiard), the statistics of energy levels in each
symmetry class should be computed separately
\end{thebibliography}
%

\printindex

\end{document}